# Performance Evaluation of Aodv&DSR with Varying Pause Time&Node Density Over TCP&CBR Connections in Vanet

**Bijan Paul† Md. Ibrahim †† and Md. Abu Naser Bikas †††**

†Dept. of Computer Science & Engineering, Shahjalal University of Science & Technology, Sylhet, Bangladesh

**Summary**
Vehicular ad hoc network is formed by cars which are called nodes; allow them to communicate with one another without using any fixed road side unit. It has some unique characteristics which make it different from other ad hoc network as well as difficult to define any exact mobility model and routing protocols because of their high mobility and changing mobility pattern. Hence performance of routing protocols can vary with the various parameters such as speed, pause time, node density and traffic scenarios. In this research paper, the performance of two on-demand routing protocols AODV & DSR has been analyzed by means of packet delivery ratio, loss packet ratio & average end-to-end delay with varying pause time and node density under TCP & CBR connection.

## 1. Introduction

VANET (vehicular adhoc network) is an autonomous & self-organizing wireless communication network. In this network the cars are called nodes which involve themselves as servers and/or clients for exchanging & sharing information. This is a new technology thus government has taken huge attention on it. There are many research projects around the world which are related with VANET such as COMCAR [1], DRIVE [2], FleetNet [3] and NoW (Network on Wheels) [4], CarTALK 2000 [5], CarNet [6].

There are several VANET applications such as Vehicle collision warning, Security distance warning, Driver assistance, Cooperative driving, Cooperative cruise control, Dissemination of road information, Internet access, Map location, Automatic parking, and Driverless vehicles.
In this paper, we have evaluated performance of AODV and DSR based on TCP and CBR connection with varying pause time and also various network parameters and measured performance metrics such as packet delivery ratio, loss packet ratio and average end-to-end delay of this two routing protocol and compared their performance. The remainder of the paper is organized as follows: Section 2 describes two unicast routing protocols AODV and DSR of VANET. Section 2 describes previous work related to performance evaluation of AODV and DSR and section 3 discusses about AODV and DSR. Section 4 describes connection types like TCP and CBR. Section 5 presents performance metrics and the network parameters. Section 6 presents our implementation. We conclude in Section 7 and section 8 for reference.

## 2. Related Work

There are several papers [7, 8, 9] related to performance evaluation of AODV and DSR. We have observed that, though there is a significant difference between TCP and CBR connection but comparison in between them is not yet analyzed. So we have focused on this two connection pattern based on different connection types. In our previous work [10], we have shown how performance varies with speed limit. In this paper we observed and analyzed our experiment with varying pause time.

## 3. Routing Protocols

An ad hoc routing protocol [11] is a convention, or standard, that controls how nodes decide which way to route packets in between computing devices in a mobile adhoc network.
There are two categories of routing protocol in VANET such as Topology based routing protocols & Position based routing protocols. Existing unicast routing protocols of VANET is not capable to meet every traffic scenarios. They have some pros and cons. We have already described it in our previous work [12]. We have selected two on demand routing protocols AODV & DSR for our simulation purpose

### 3.1 AODV

Ad Hoc on Demand Distance Vector routing protocol [13] is a reactive routing protocol which establish a route when a node requires sending data packets. It has the ability of unicast & multicast routing. It uses a destination sequence number (DestSeqNum) which makes it different from other on demand routing protocols. It maintains routing tables, one entry per destination and an entry is discarded if it is not used recently. It establishes route by using

RREQ and RREP cycle. If any link failure occurs, it sends report and another RREQ is made.

### 3.2 DSR

The Dynamic Source Routing (DSR) [14] protocol utilizes source routing & maintains active routes. It has two phases route discovery & route maintenance. It does not use periodic routing message. It will generate an error message if there is any link failure. All the intermediate nodes ID are stored in the packet header of DSR. If there has multiple paths to go to the destination DSR stores multiple path of its routing information.

AODV and DSR have some significant differences. In AODV when a node sends a packet to the destination then data packets only contains destination address. On the other hand in DSR when a node sends a packet to the destination the full routing information is carried by data packets which causes more routing overhead than AODV.

## 4. Connection Types

There are several types of connection pattern in VANET. For our simulation purpose we have used CBR and TCP connection pattern.

### 4.1 Constant Bit Rate (CBR)

Constant bit rate means consistent bits rate in traffic are supplied to the network. In CBR, data packets are sent with fixed size and fixed interval between each data packets. Establishment phase of connection between nodes is not required here, even the receiving node don’t send any acknowledgement messages. Connection is one way direction like source to destination.

### 4.2 Transmission Control Protocol (TCP)

TCP is a connection oriented and reliable transport protocol. To ensure reliable data transfer TCP uses acknowledgement, time outs and retransmission. Acknowledge means successful transmission of packets from source to destination. If an acknowledgement is not received during a certain period of time which is called time out then TCP transmit the data again.

## 5. Performance Metrics & Network Parameters

For network simulation, there are several performance metrics which is used to evaluate the performance. In simulation purpose we have used three performance metrics.

### 5.1 Packet Delivery Ratio

Packet delivery ratio is the ratio of number of packets received at the destination to the number of packets sent from the source. The performance is better when packet delivery ratio is high.

### 5.2 Average end-to-end delay

This is the average time delay for data packets from the source node to the destination node. To find out the end-to-end delay the difference of packet sent and received time was stored and then dividing the total time difference over the total number of packet received gave the average end-to-end delay for the received packets. The performance is better when packet end-to-end delay is low.

### 5.3 Loss Packet Ratio (LPR)

Loss Packet Ratio is the ratio of the number of packets that never reached the destination to the number of packets originated by the source.

## 6. Our Implementation

For simulation purpose we used random waypoint mobility model. Network Simulator NS-2.34[15, 16] has been used. To measure the performance of AODV and DSR we used same scenario for both protocols. Because of both protocols unique behavior the resultant output differ.

### 6.1 Simulation Parameters

In our simulation, we used environment size 840 m x 840 m, node density 30 to 150 nodes with constant maximum speed 15 m/s and variable pause time 50 to 250 s. We did the Simulation for 200s with maximum 8 connections. The network parameters we have used for our simulation purpose shown in the table 1.

Table 1 .Network Parameters

| ***Parameter*** | ***Value*** |
|---|---|
| Protocols | AODV, DSR |
| Simulation Time | 200 s |
| Number of Nodes | 30, 60, 90, 120, 150 |
| Simulation Area | 840 m x 840 m |
| Pause Time | 50,100,150,200,250s |
| Traffic Type | CBR , TCP |
| Maximum Speed | 15 m / s |
| Mobility Model | Random Waypoint |
| Network Simulator | NS 2.34 |

## 6.2 Simulation Results

The performance of AODV & DSR has been analyzed with varying pause time 50s to 250s for number of nodes 30, 60, 90, 120, 150 under TCP & CBR connection. We measure the packet delivery ratio, loss packet ratio & average end-to-end delay of AODV and DSR and the simulated output has shown by using graphs.

## 6.3 Graphs

On left side in module 5.3 we draw the graph of TCP connection simulation result. Similarly, on right side we draw the graph of CBR connection simulation result.

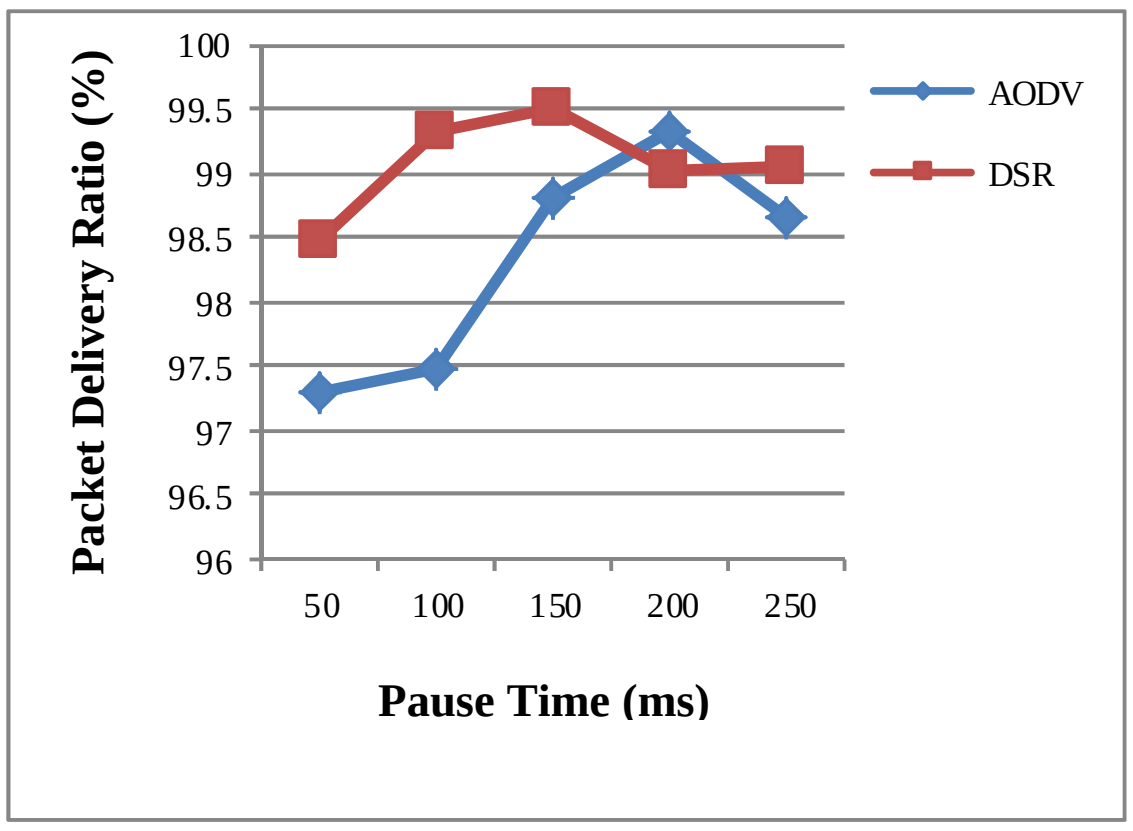

Fig 1: PDR of 30 nodes using TCP

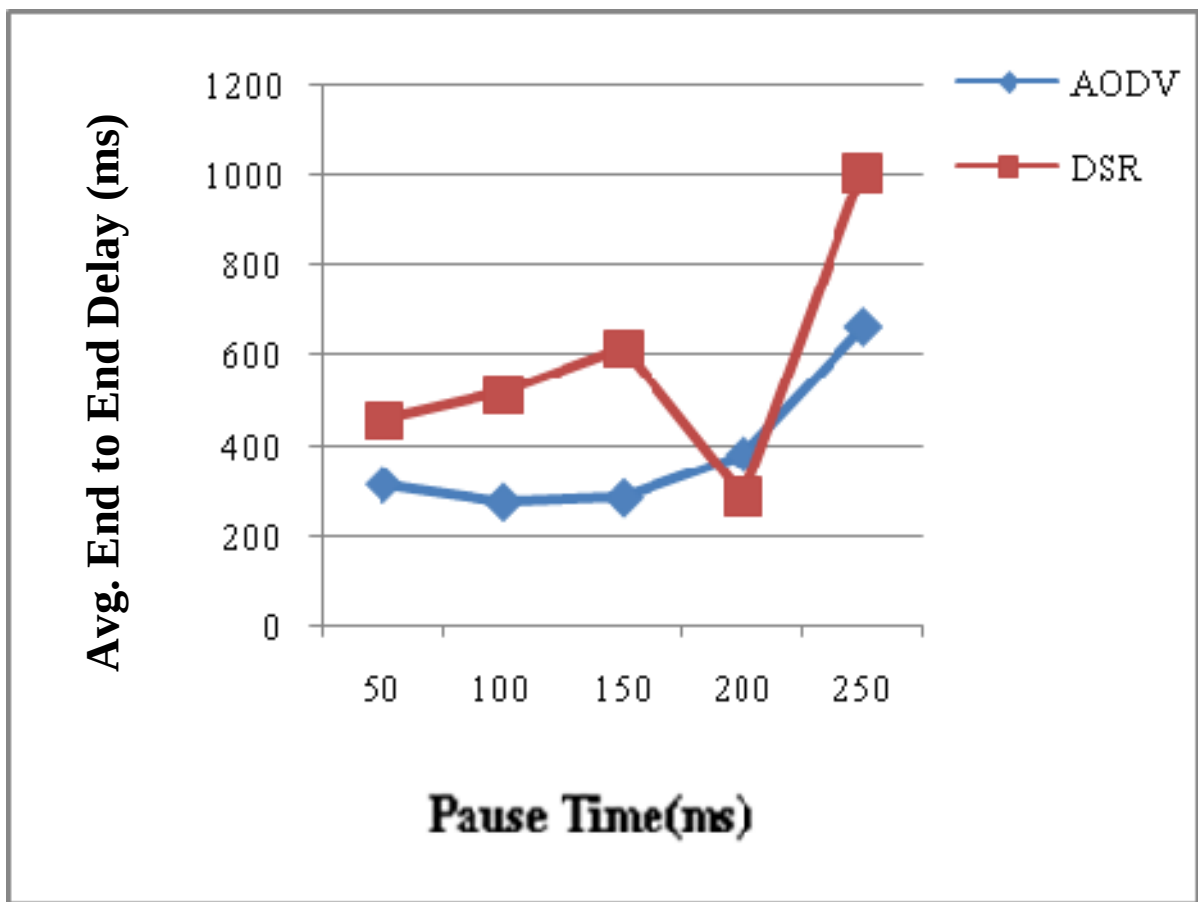

Fig 2: Avg.E-2-E delay of 30 nodes using TCP

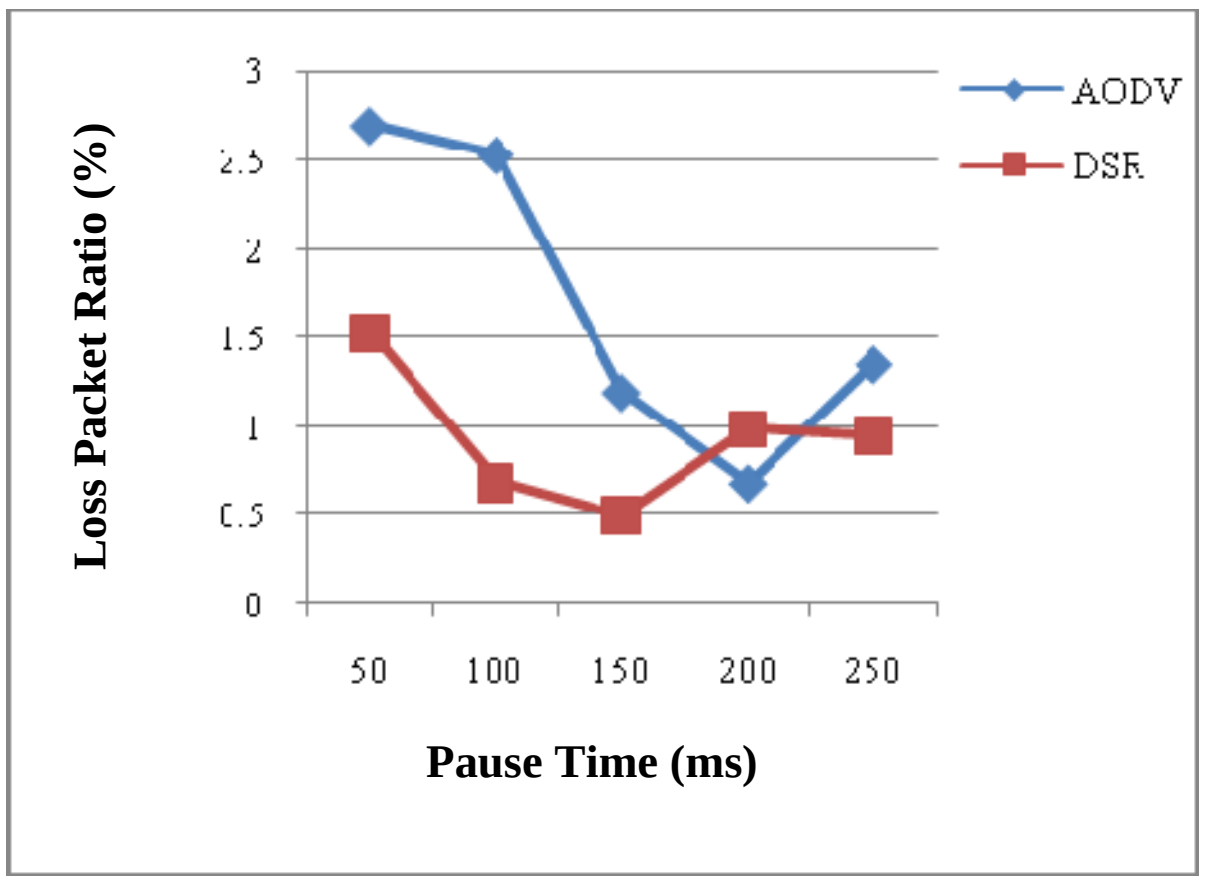

Fig 3: LPR of 30 nodes using TCP

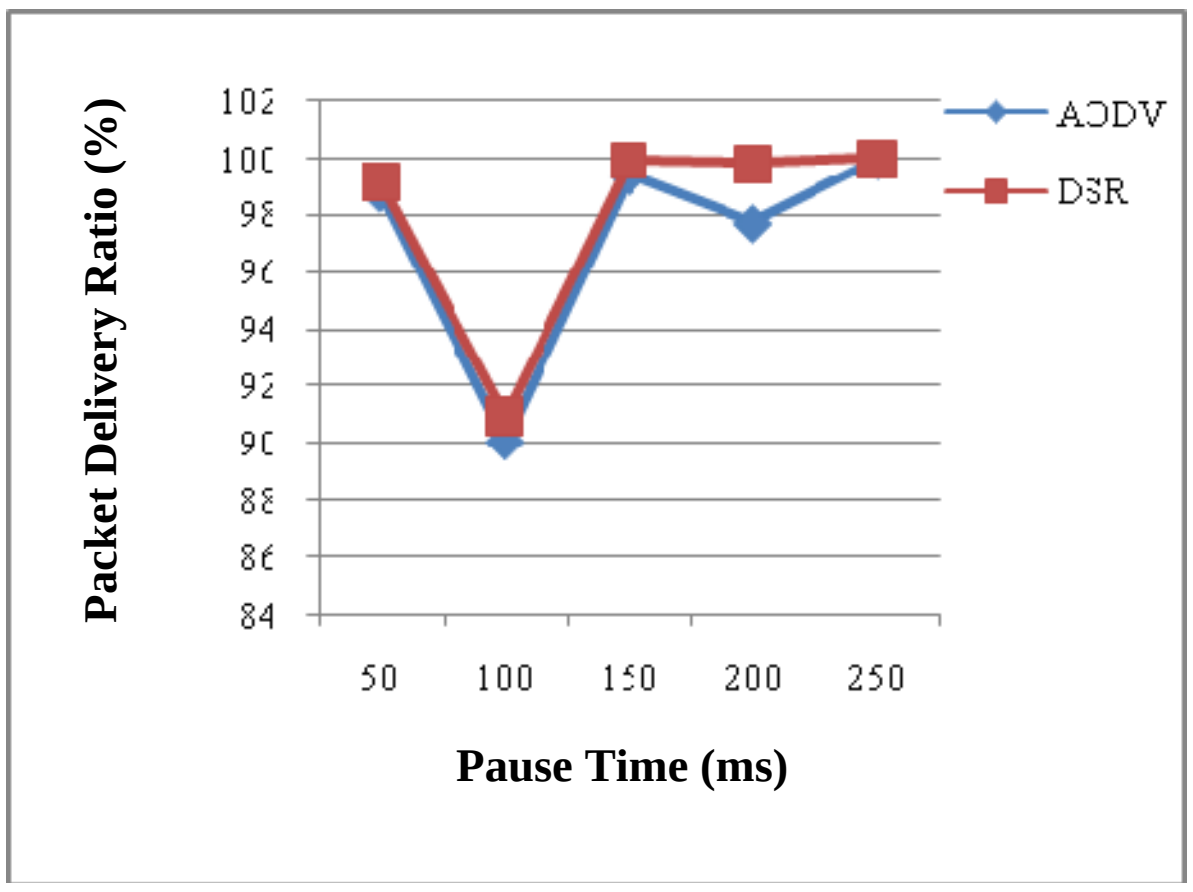

Fig 4: PDR of 30 nodes using CBR

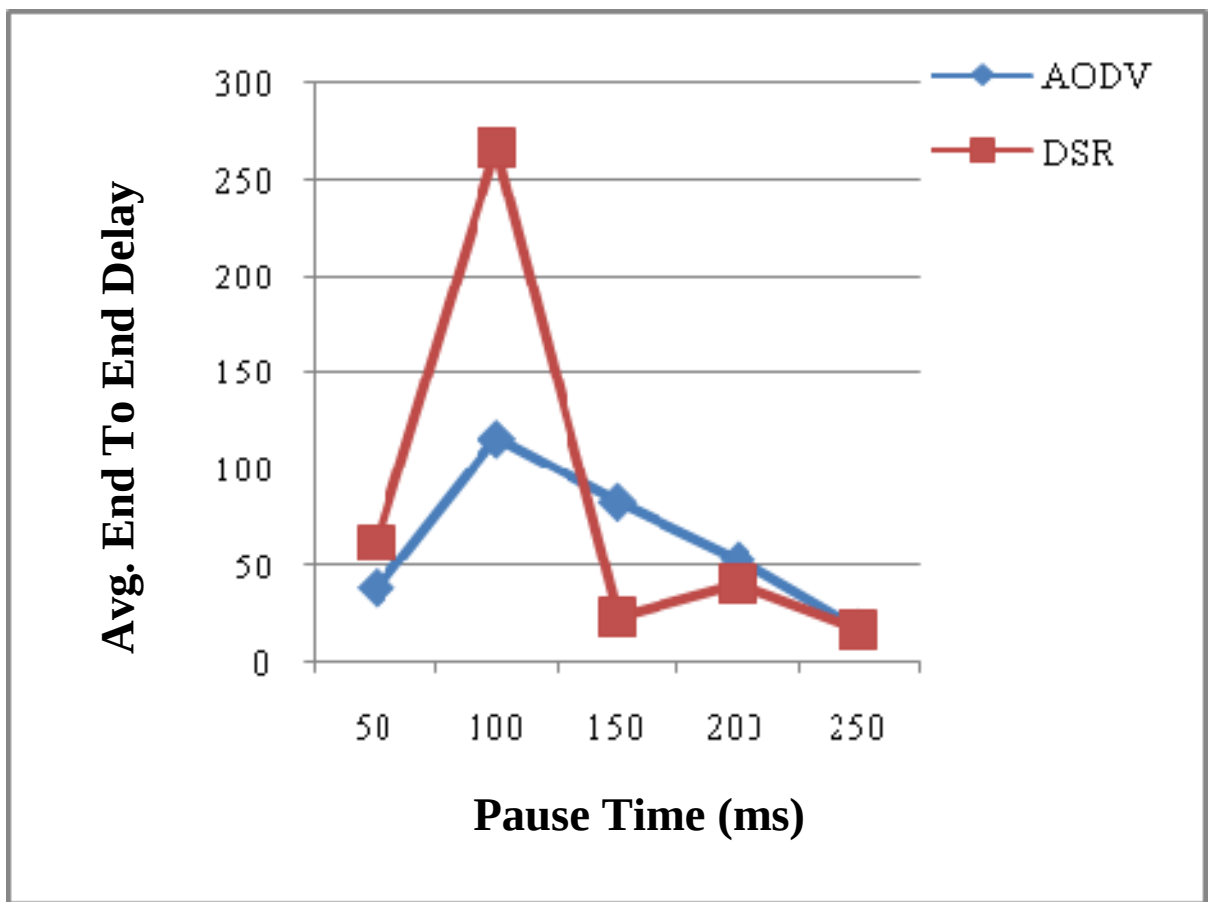

Fig 5: Avg.E-2-E delay of 30 nodes using CBR

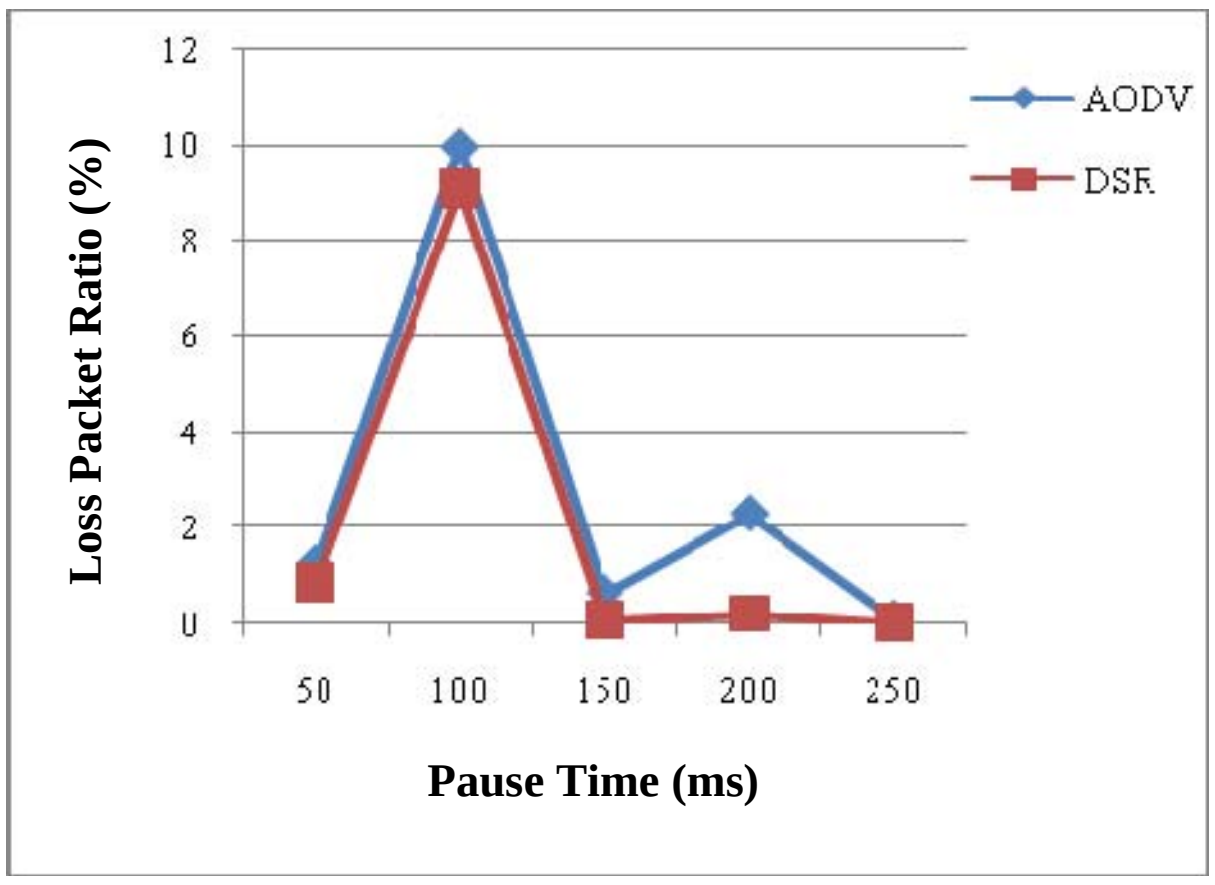

Fig 6: LPR of 30 nodes using CBR

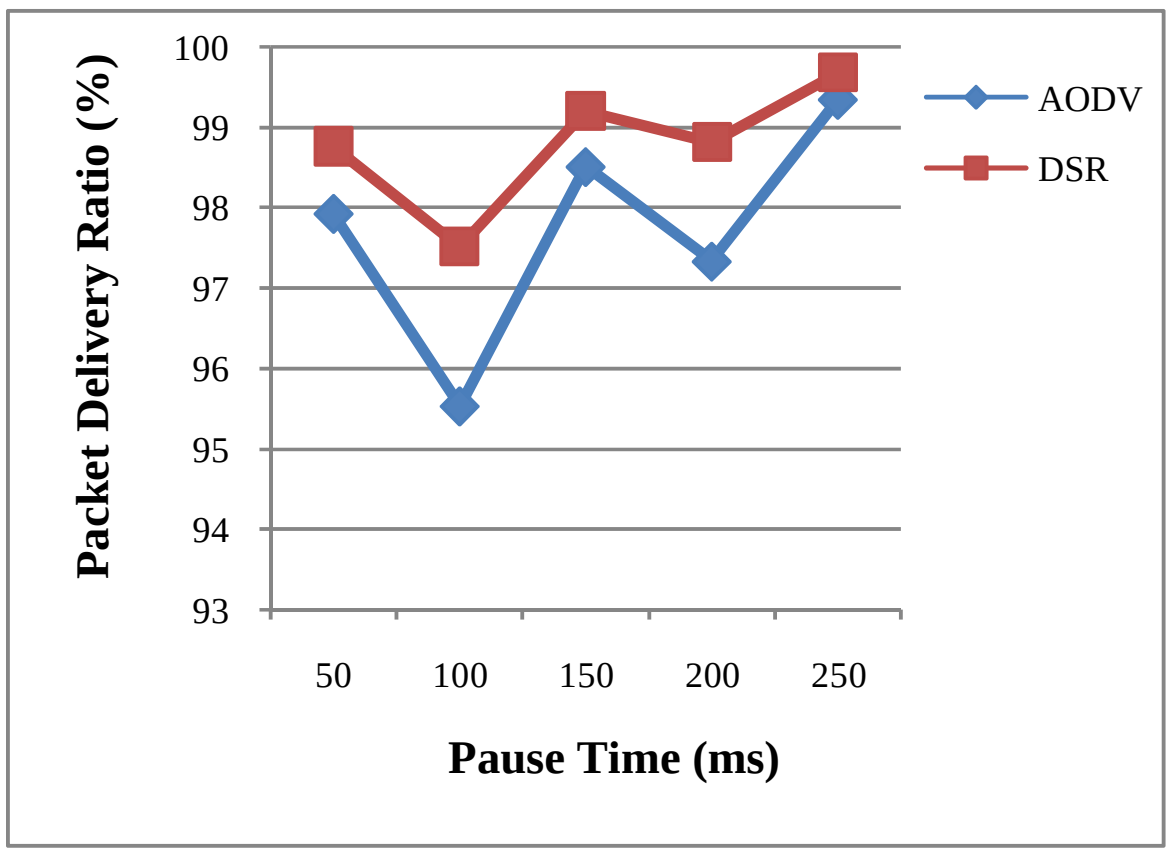

Fig 7: PDR of 60 nodes using TCP

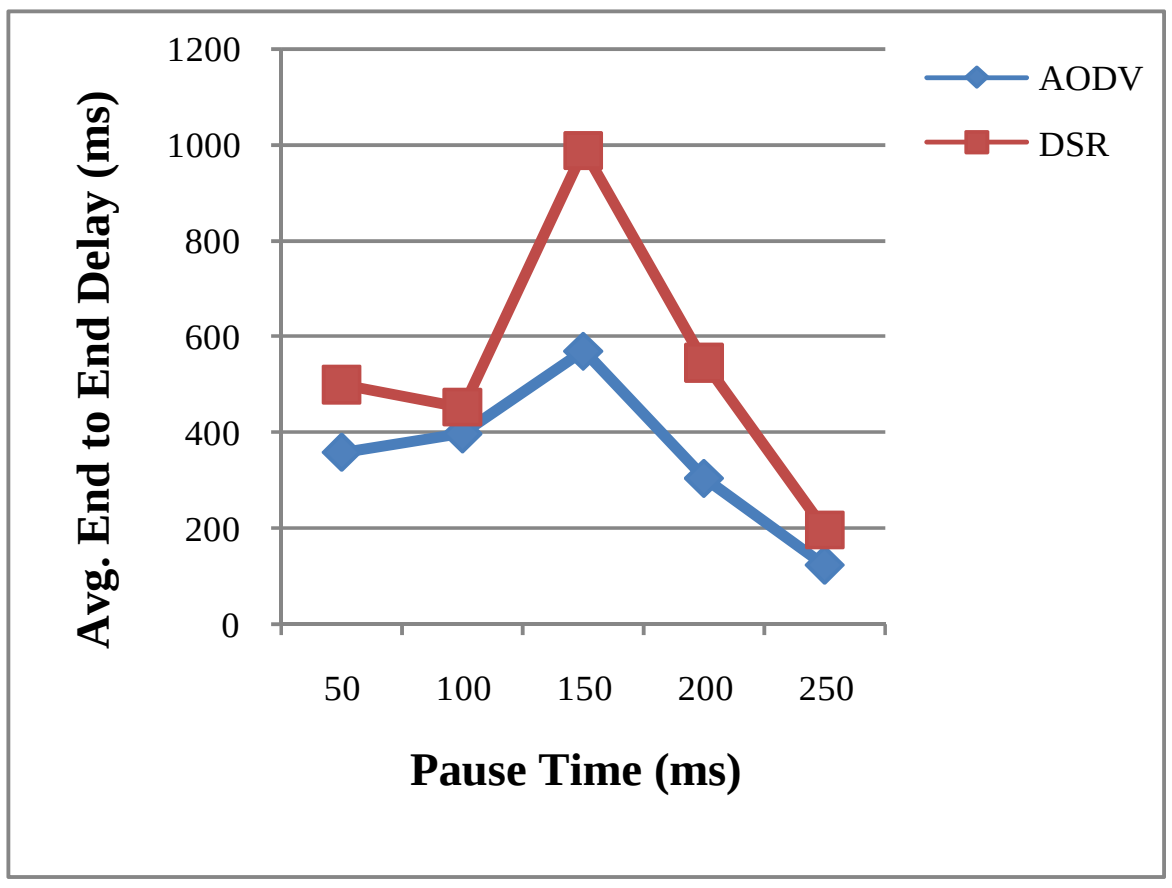

Fig 8: Avg.E-2-E delay of 60 nodes using TCP

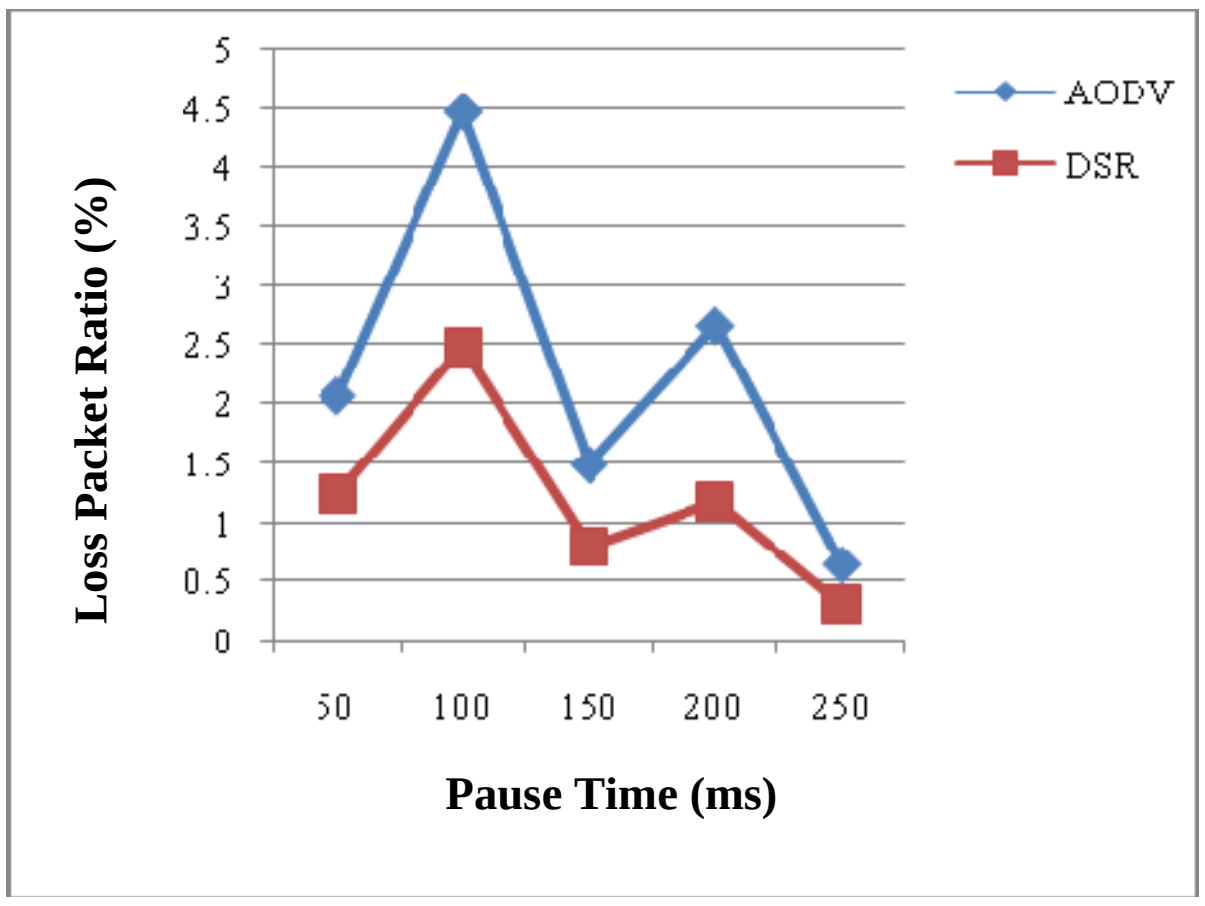

Fig 9: LPR of 60 nodes using TCP

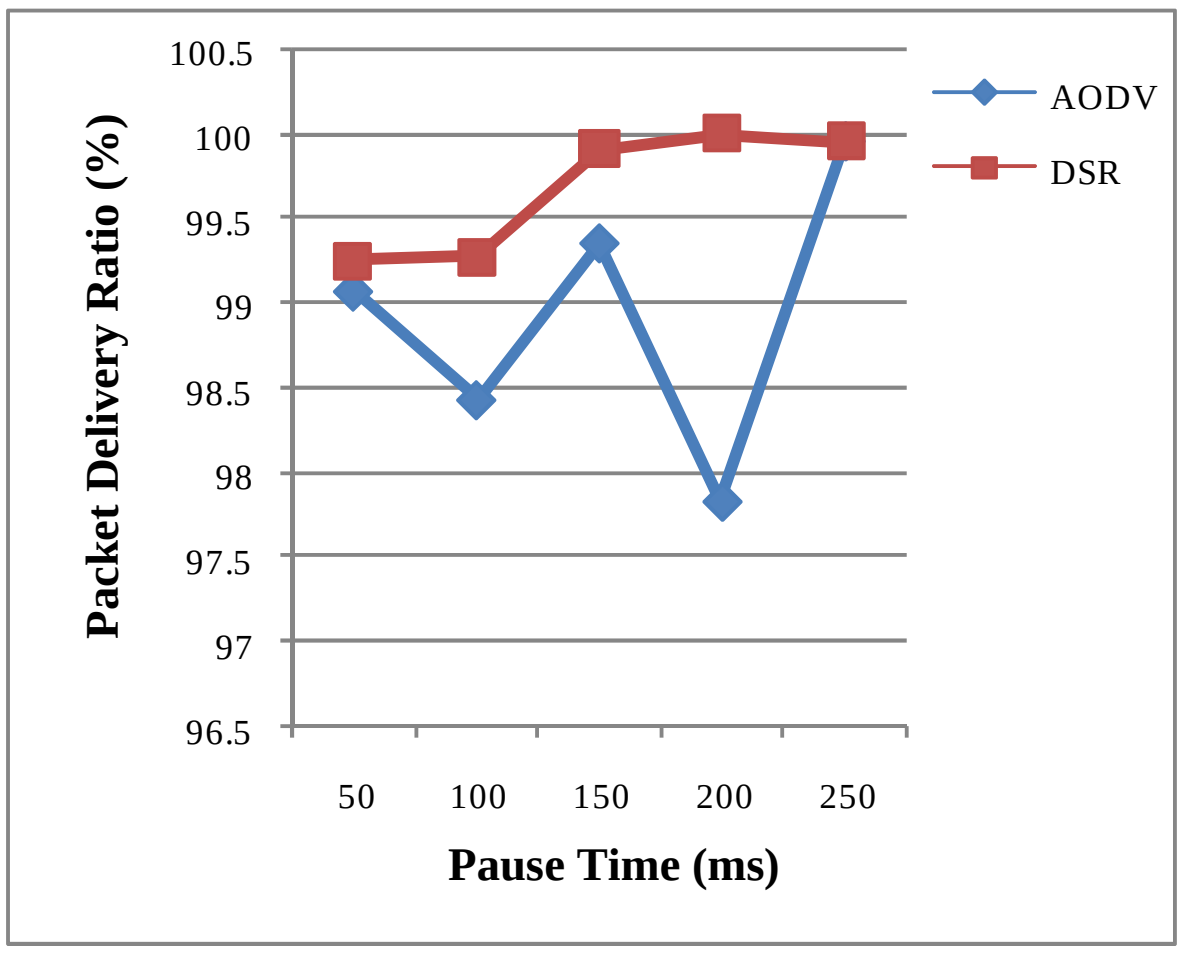

Fig 10: PDR of 60 nodes using CBR

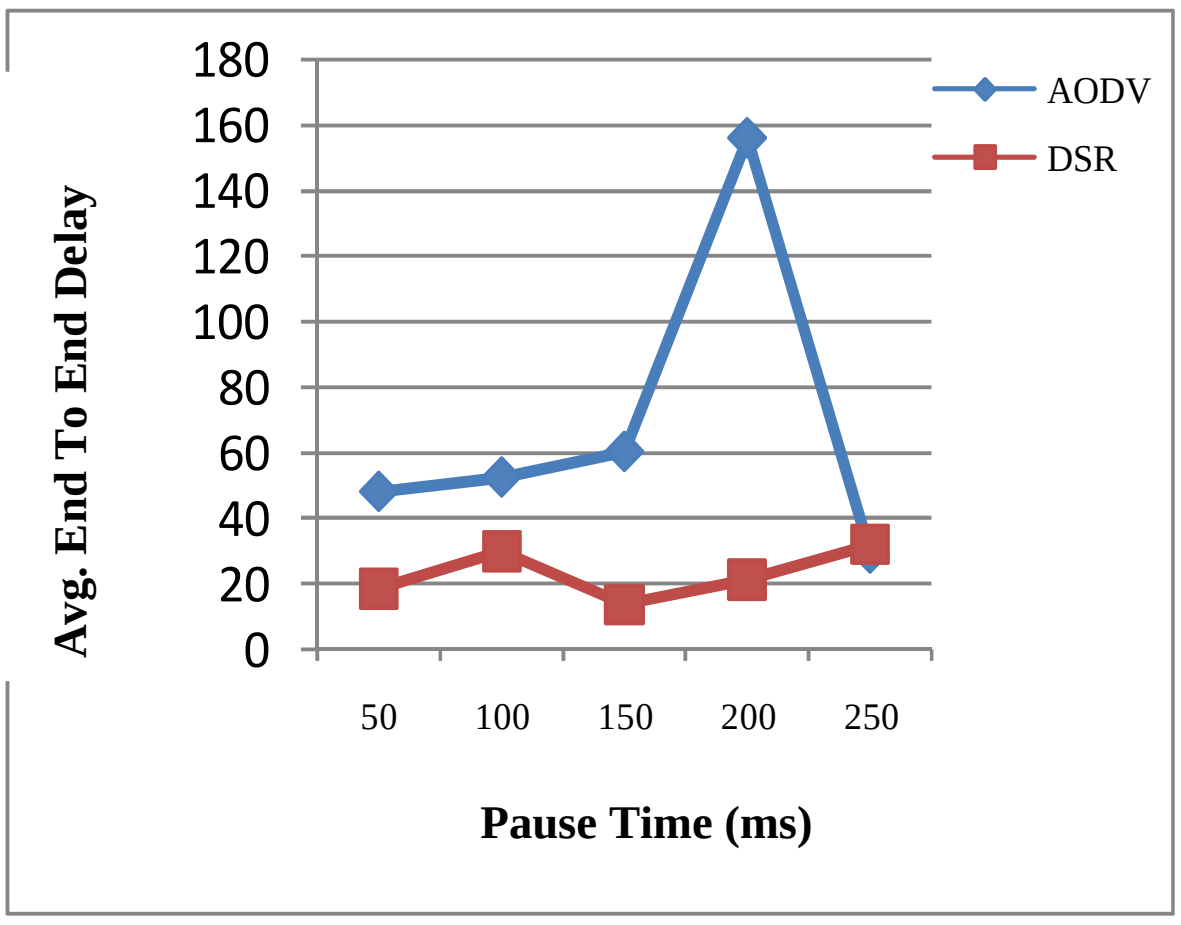

Fig 11: Avg.E-2-E delay of 60 nodes using CBR

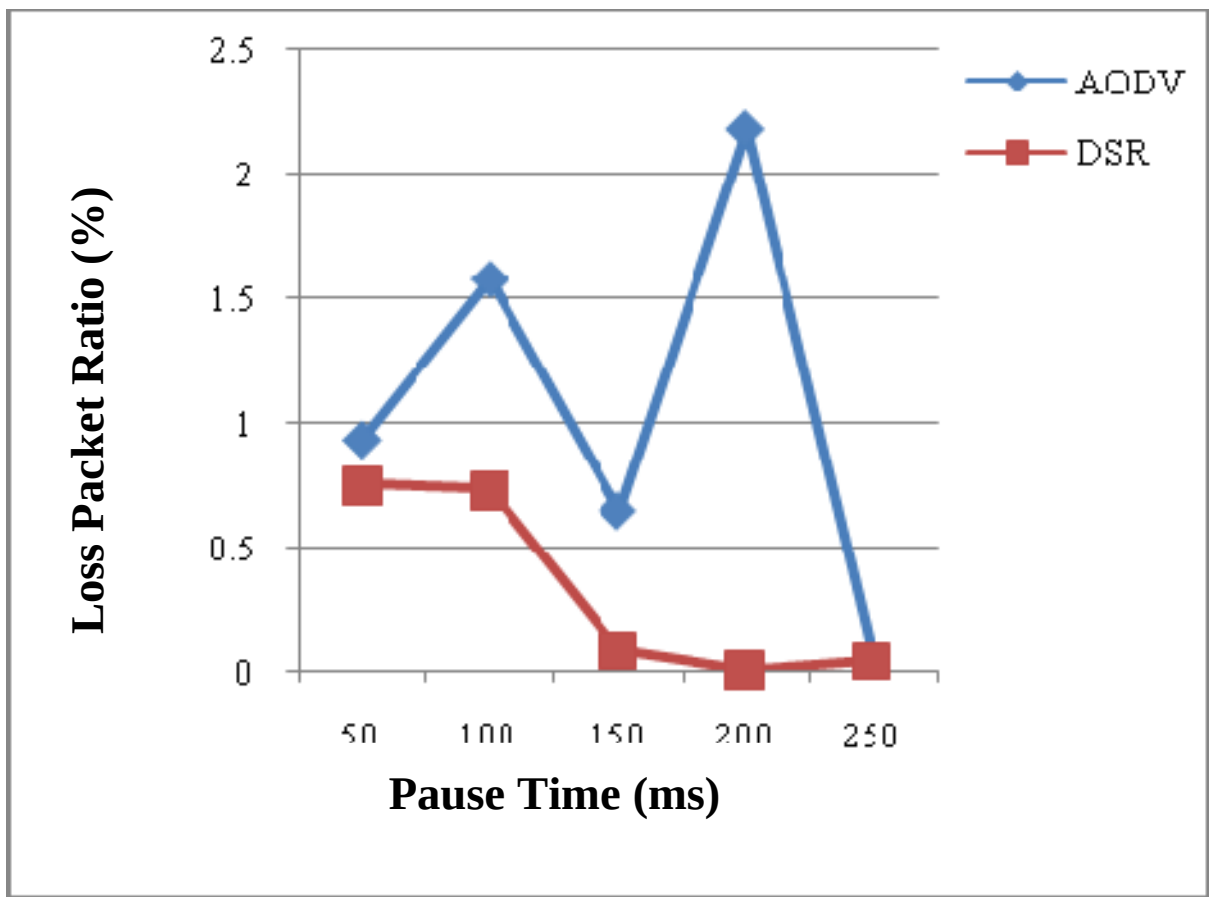

Fig 12: LPR of 60 nodes using CBR

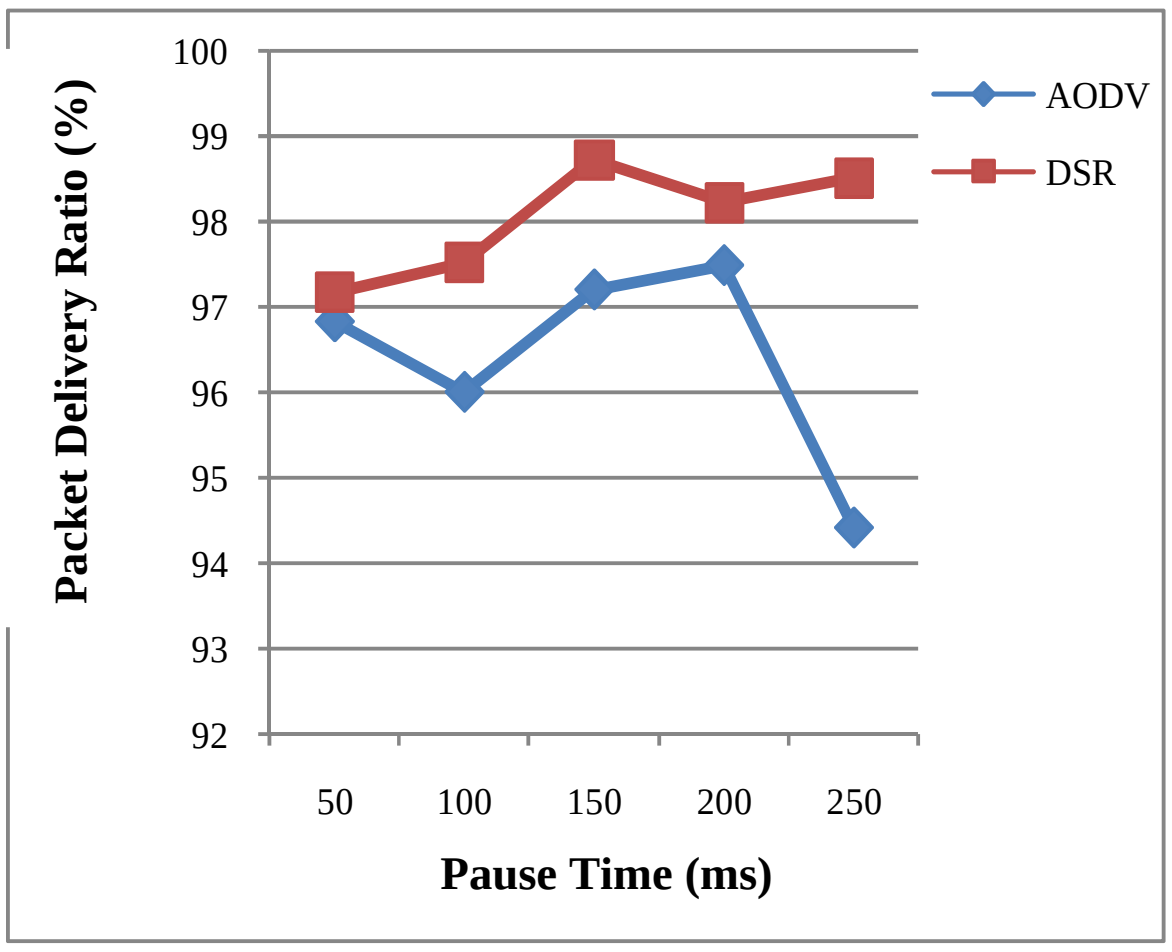

Fig 13: PDR of 90 nodes using TCP

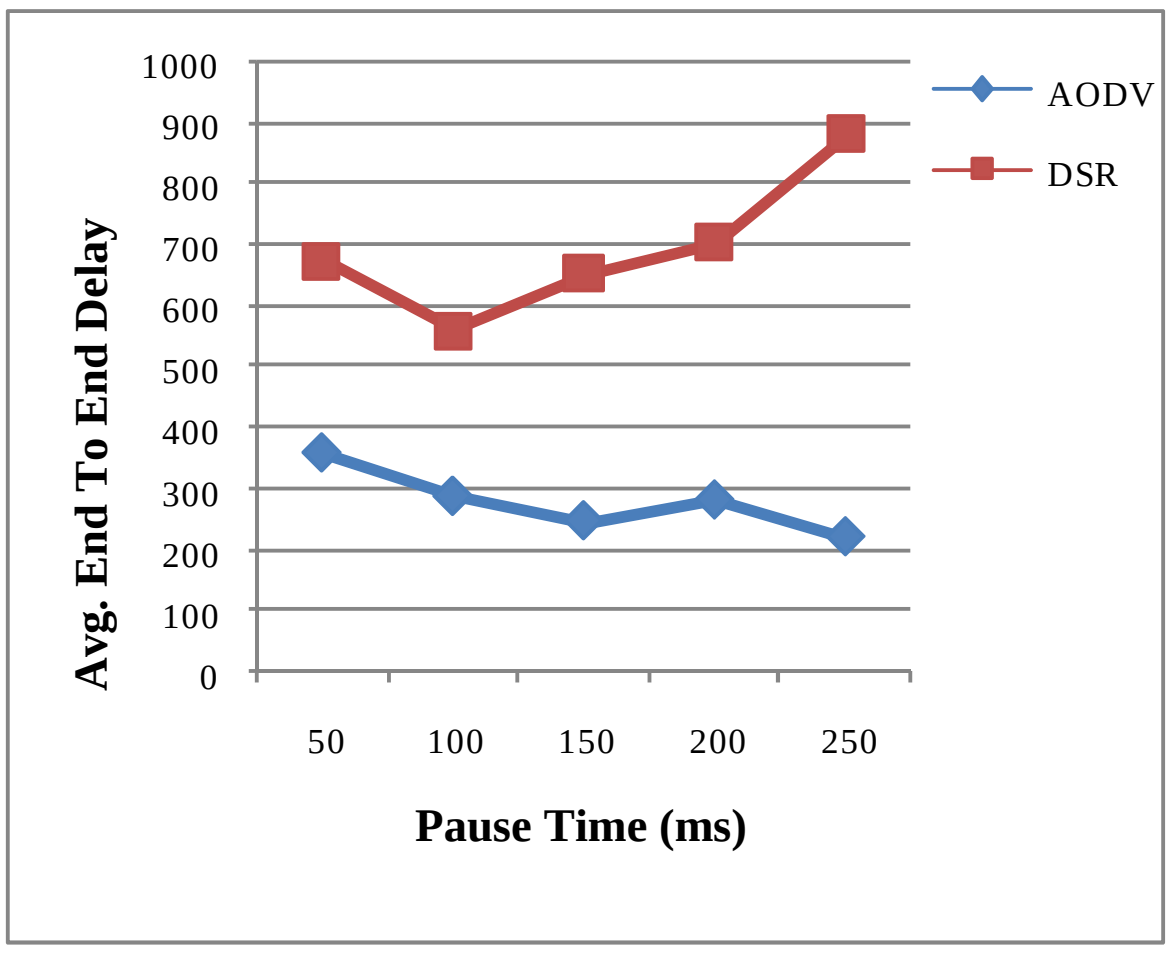

Fig 14: Avg.E-2-E delay of 90 nodes using TCP

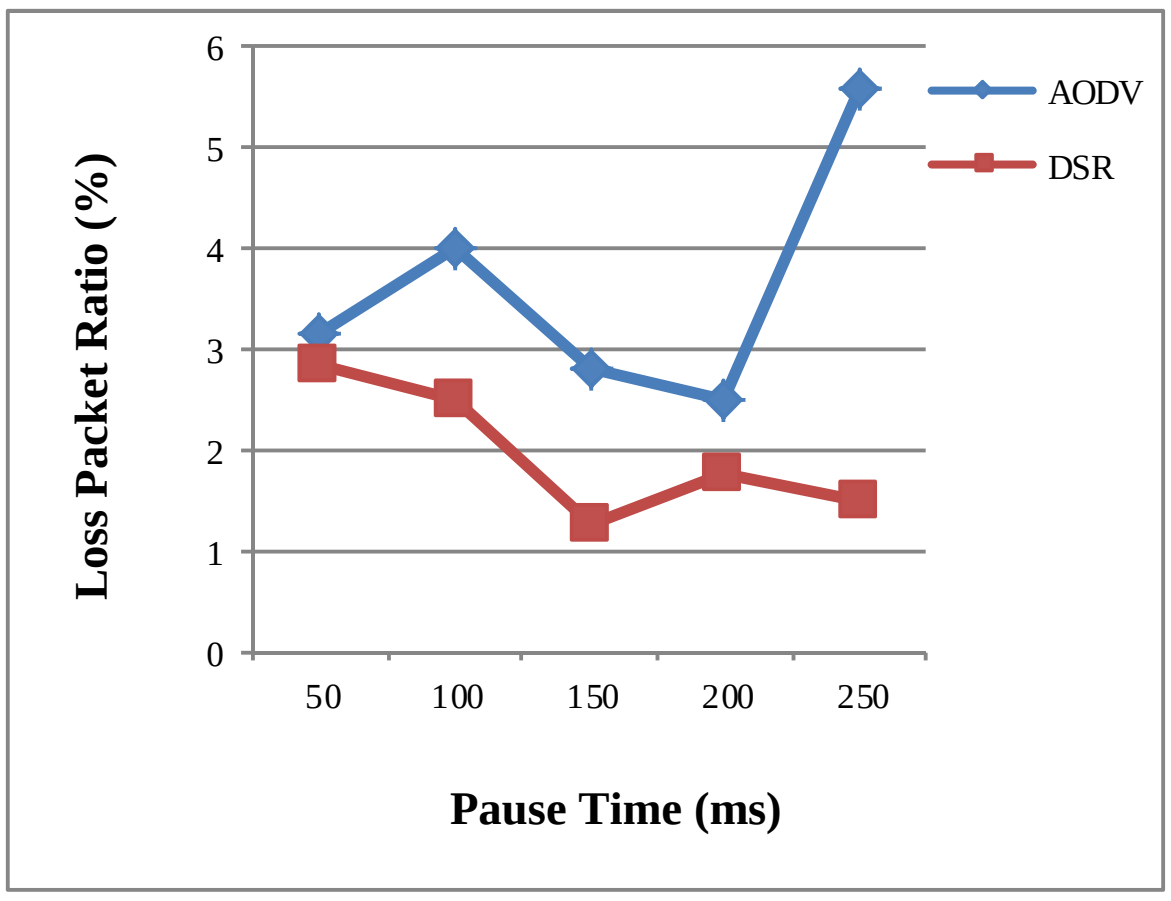

Fig 15: LPR of 90 nodes using TCP

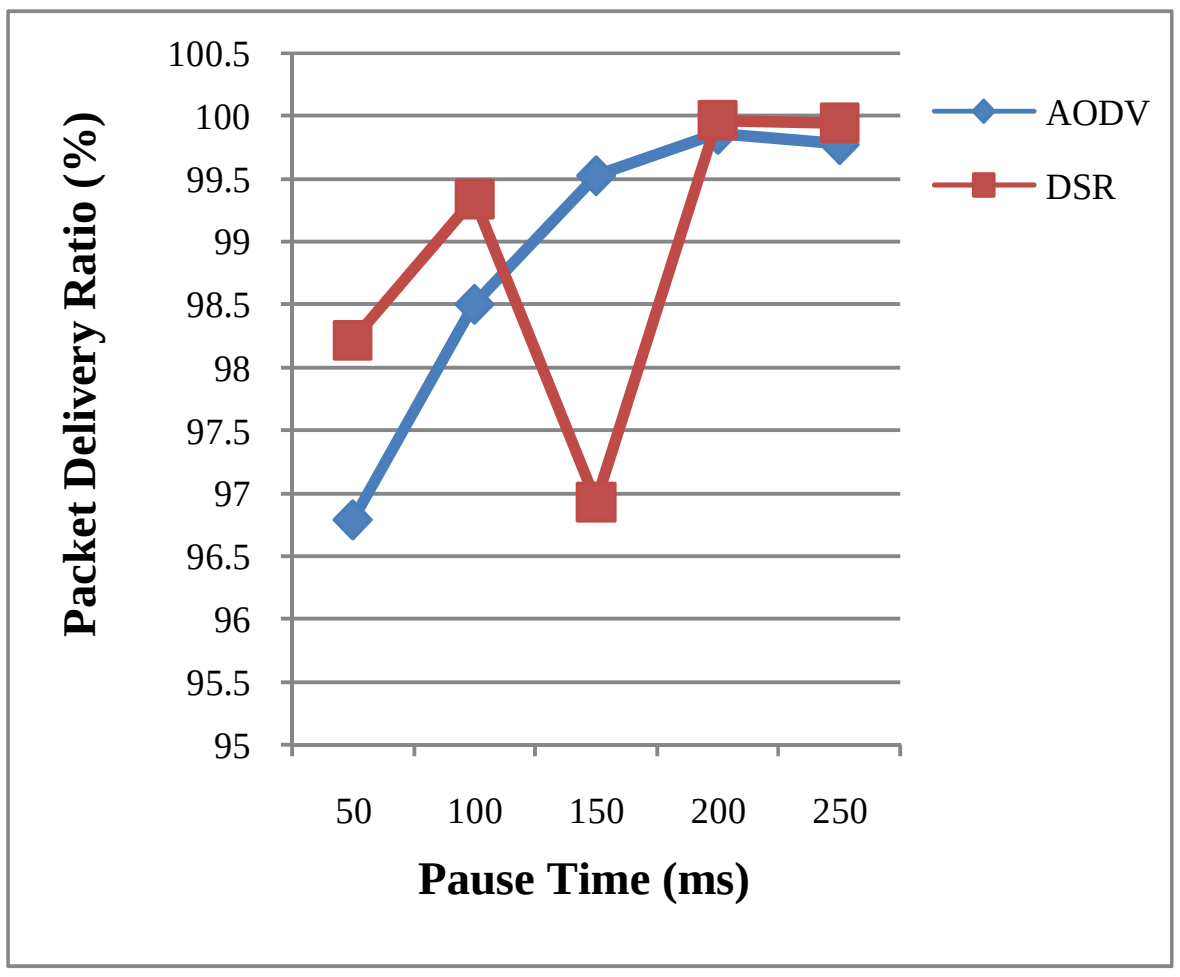

Fig 16: PDR of 90 nodes using CBR

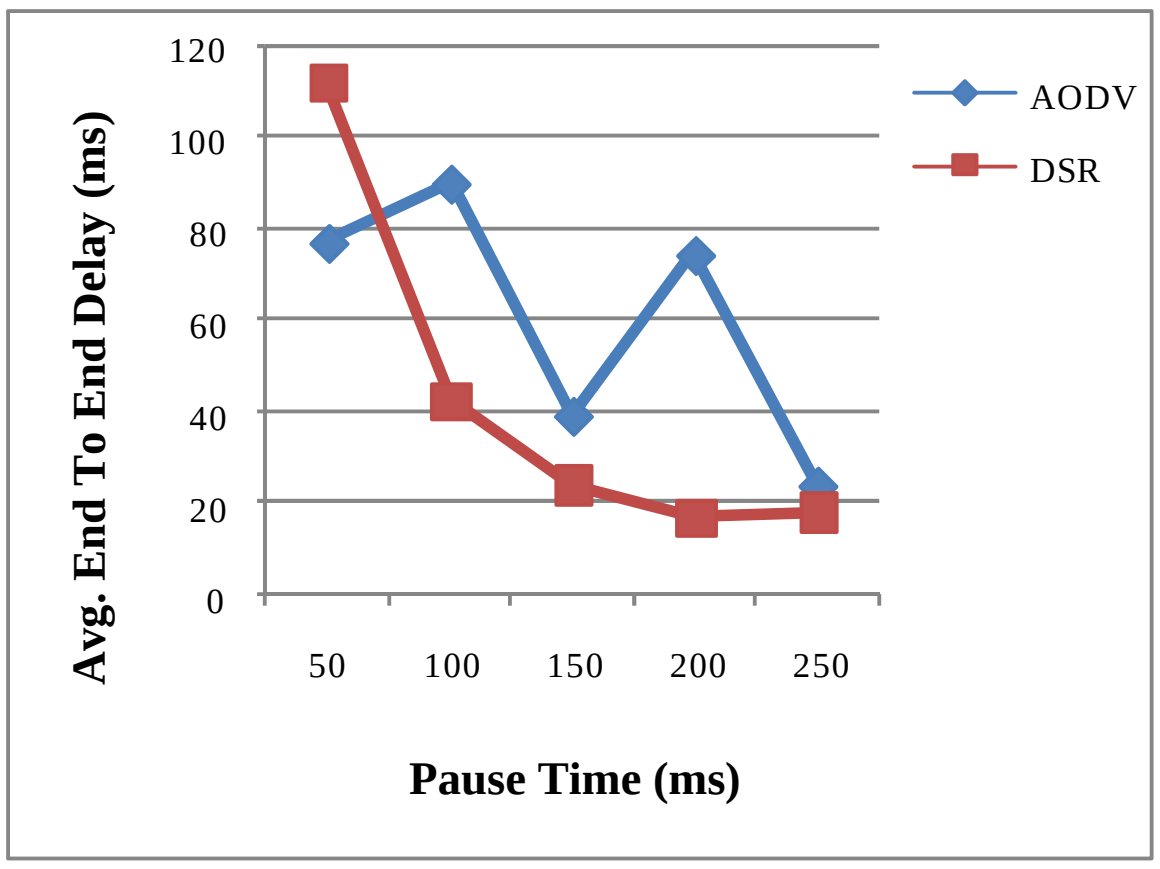

Fig 17: Avg.E-2-E delay of 90 nodes using CBR

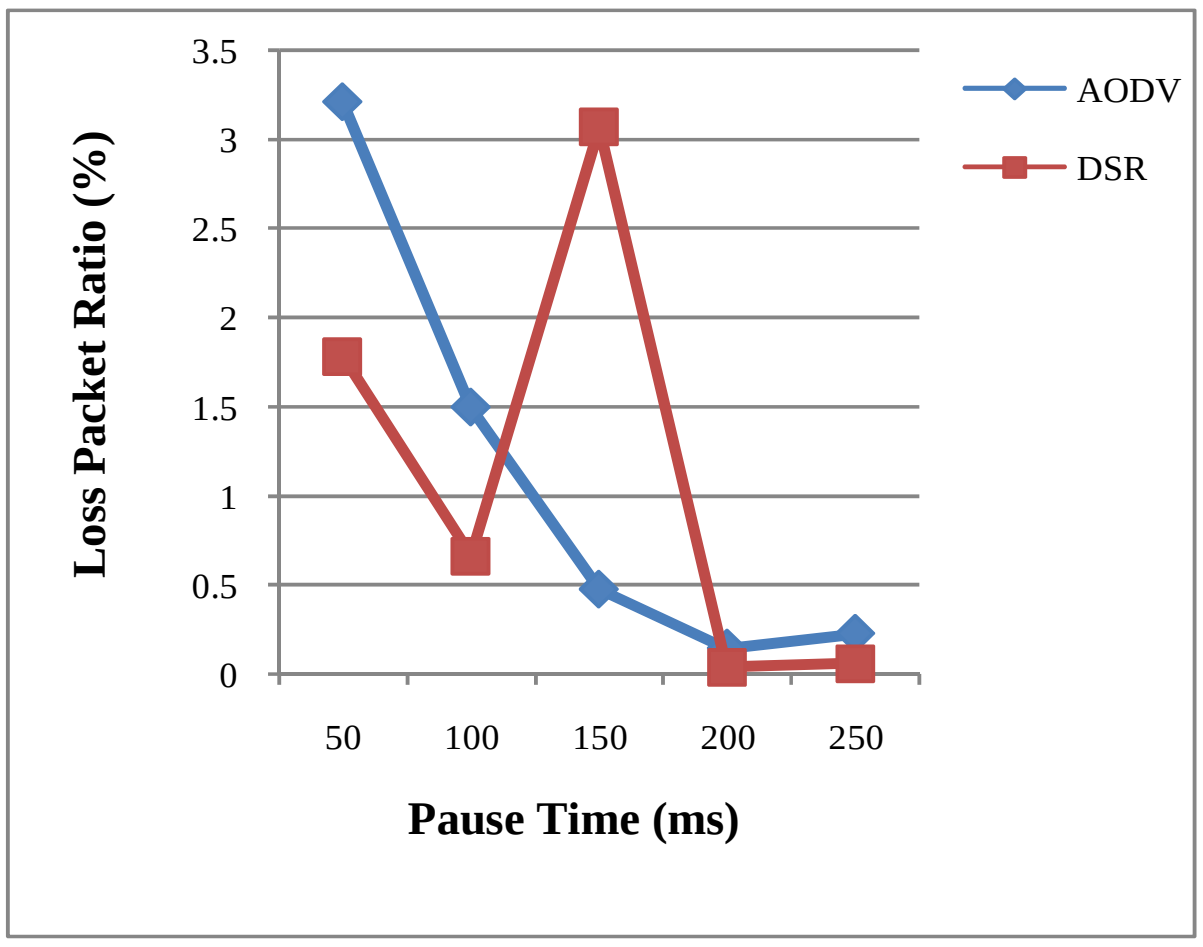

Fig 18: LPR of 90 nodes using CBR

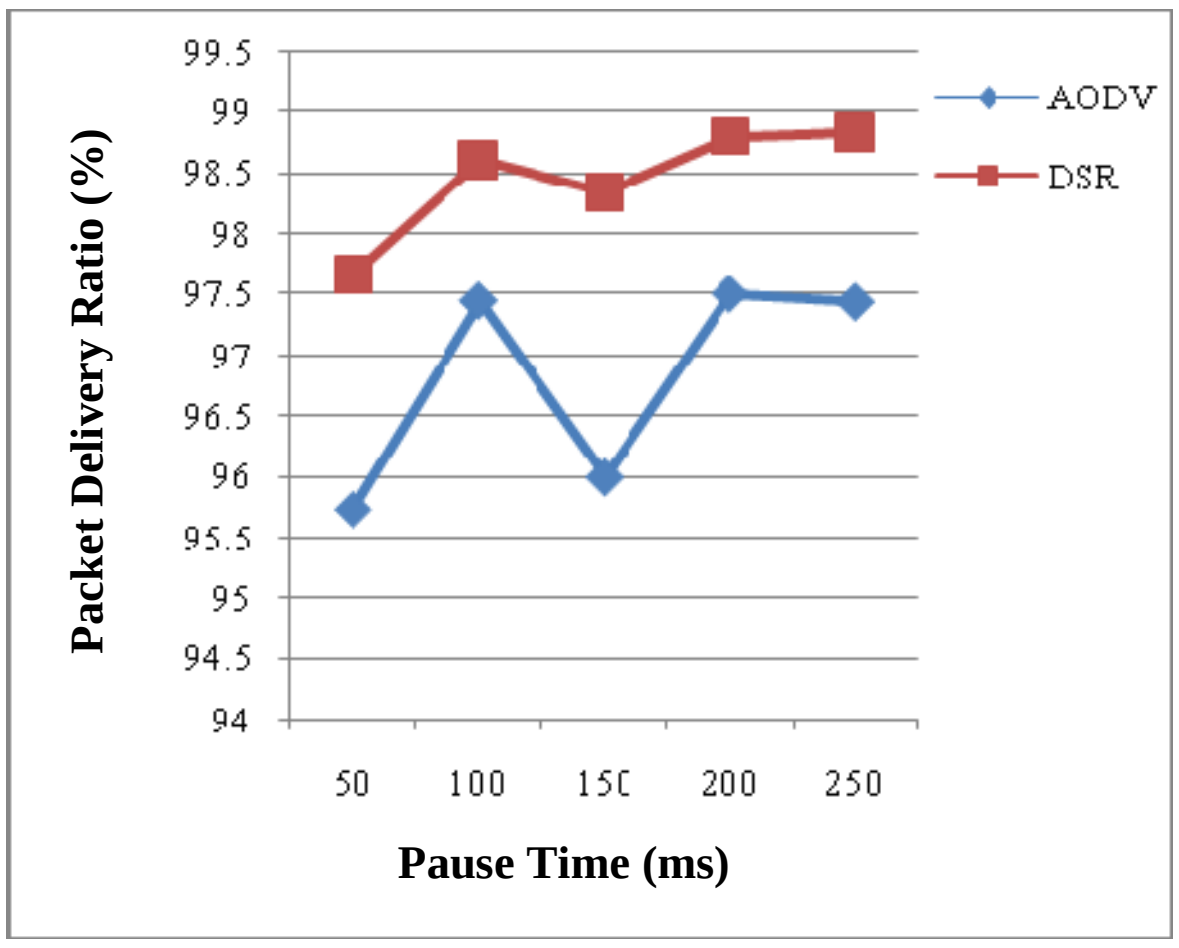

Fig 19: PDR of 120 nodes using TCP

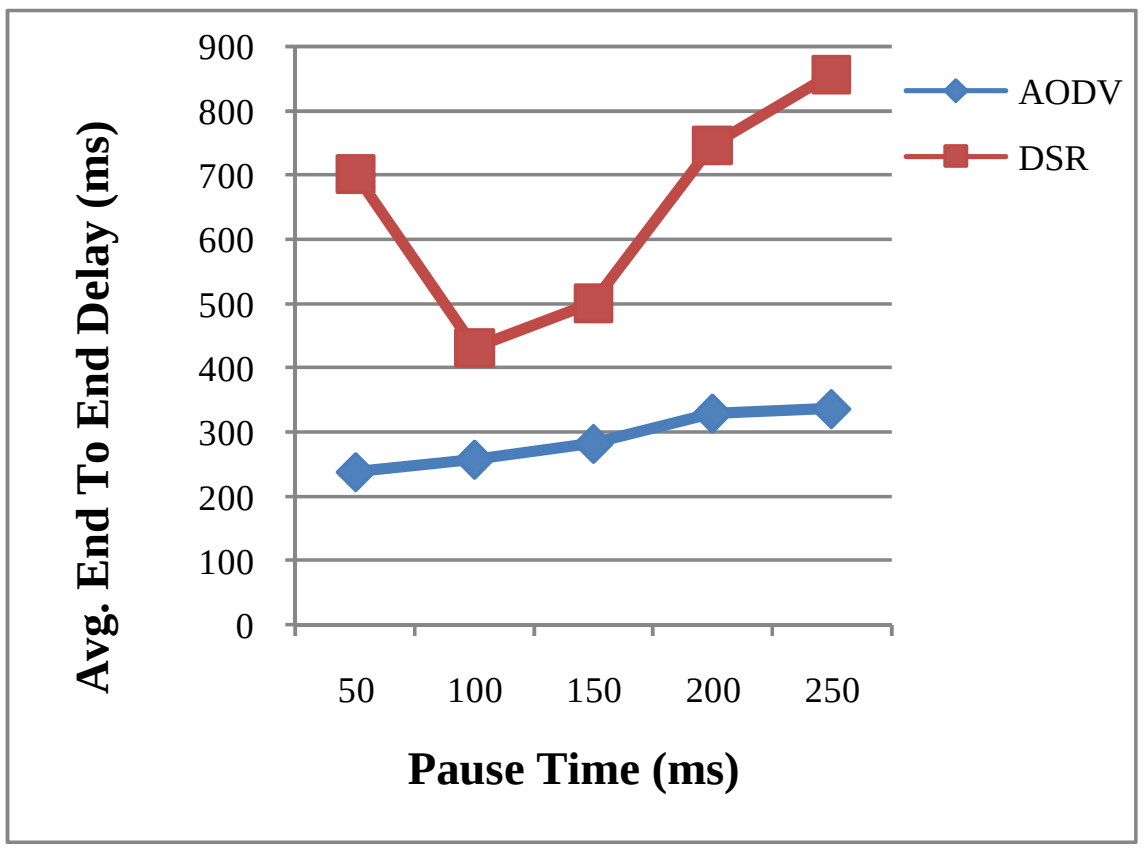

Fig 20: Avg.E-2-E delay of 120 nodes using TCP

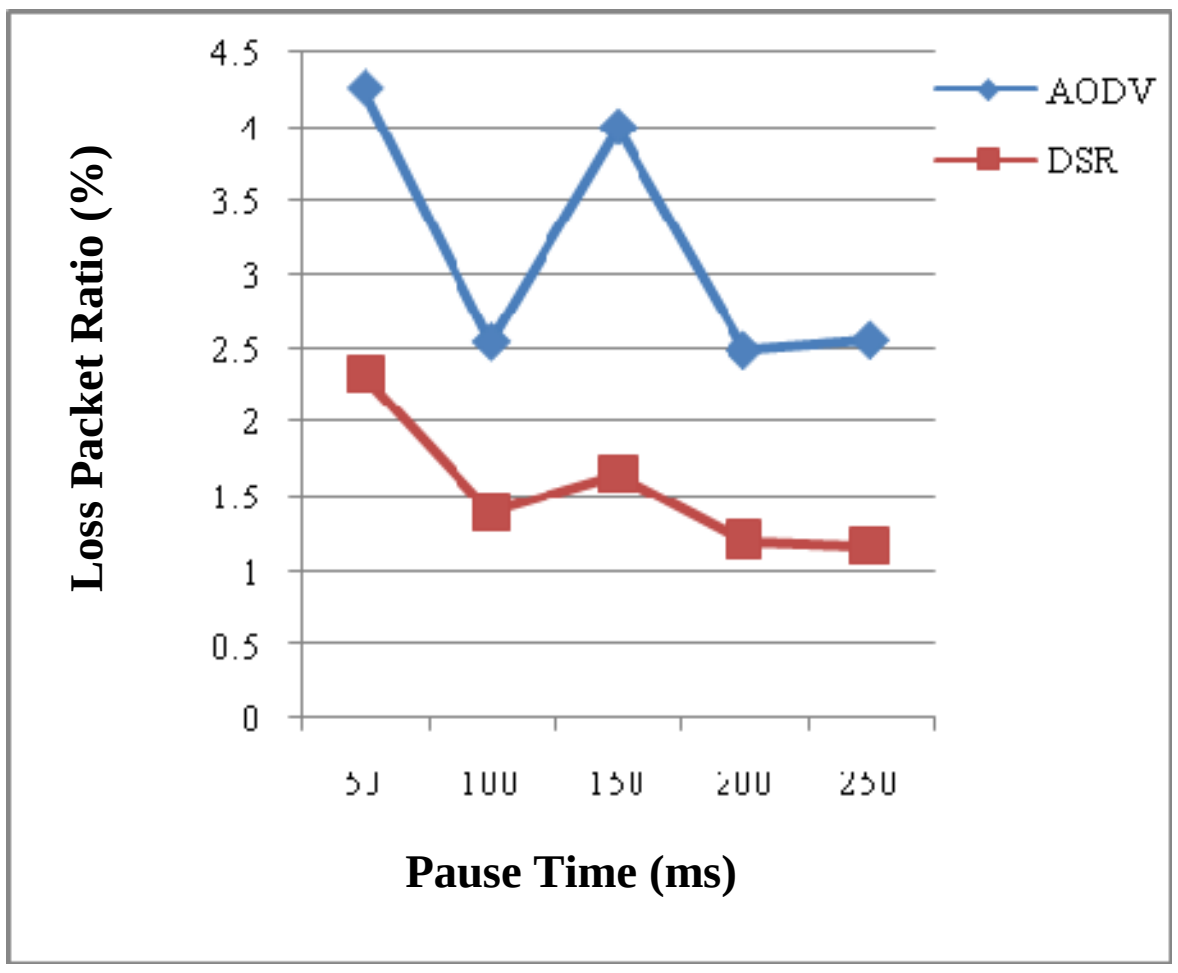

Fig 21: LPR of 120 nodes using TCP

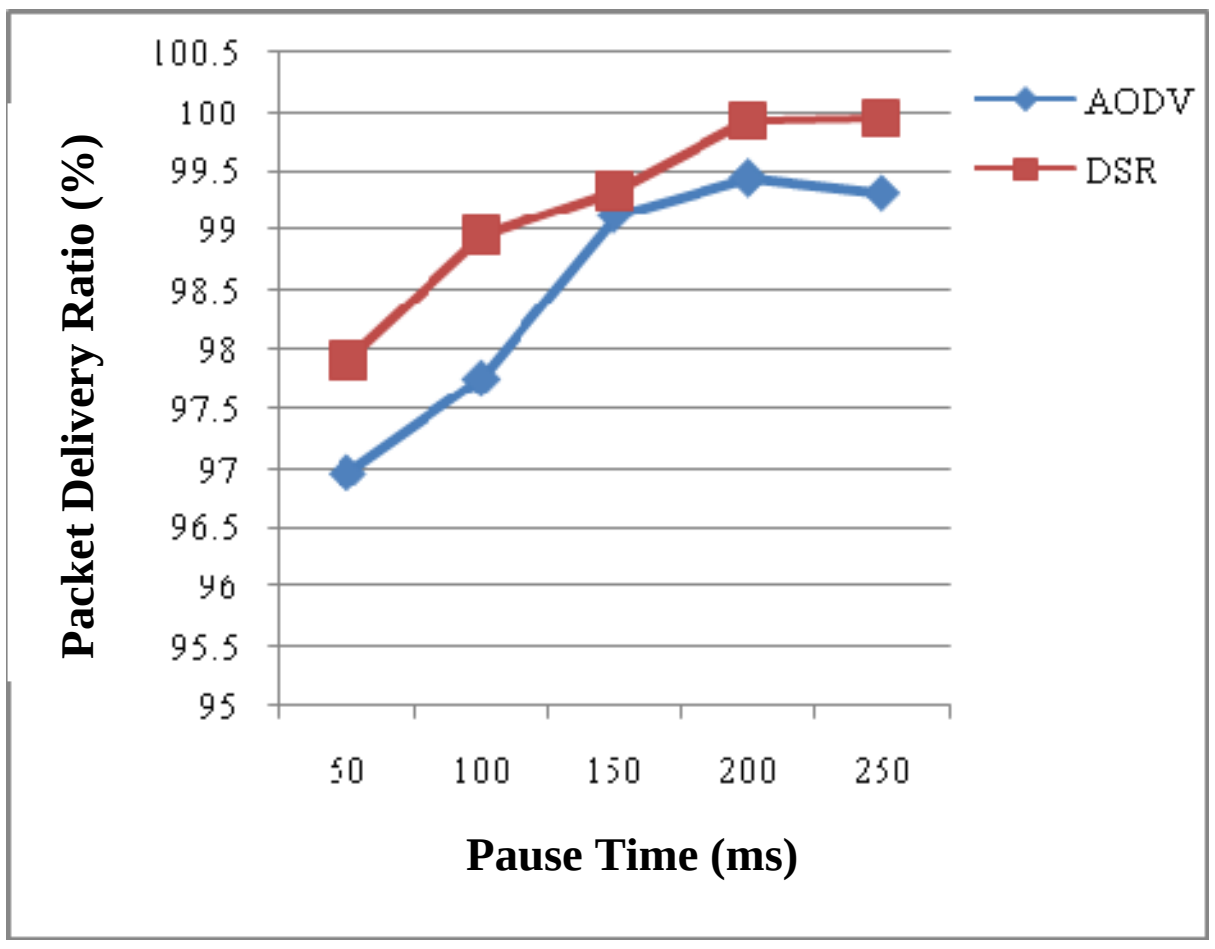

Fig 22: PDR of 120 nodes using CBR

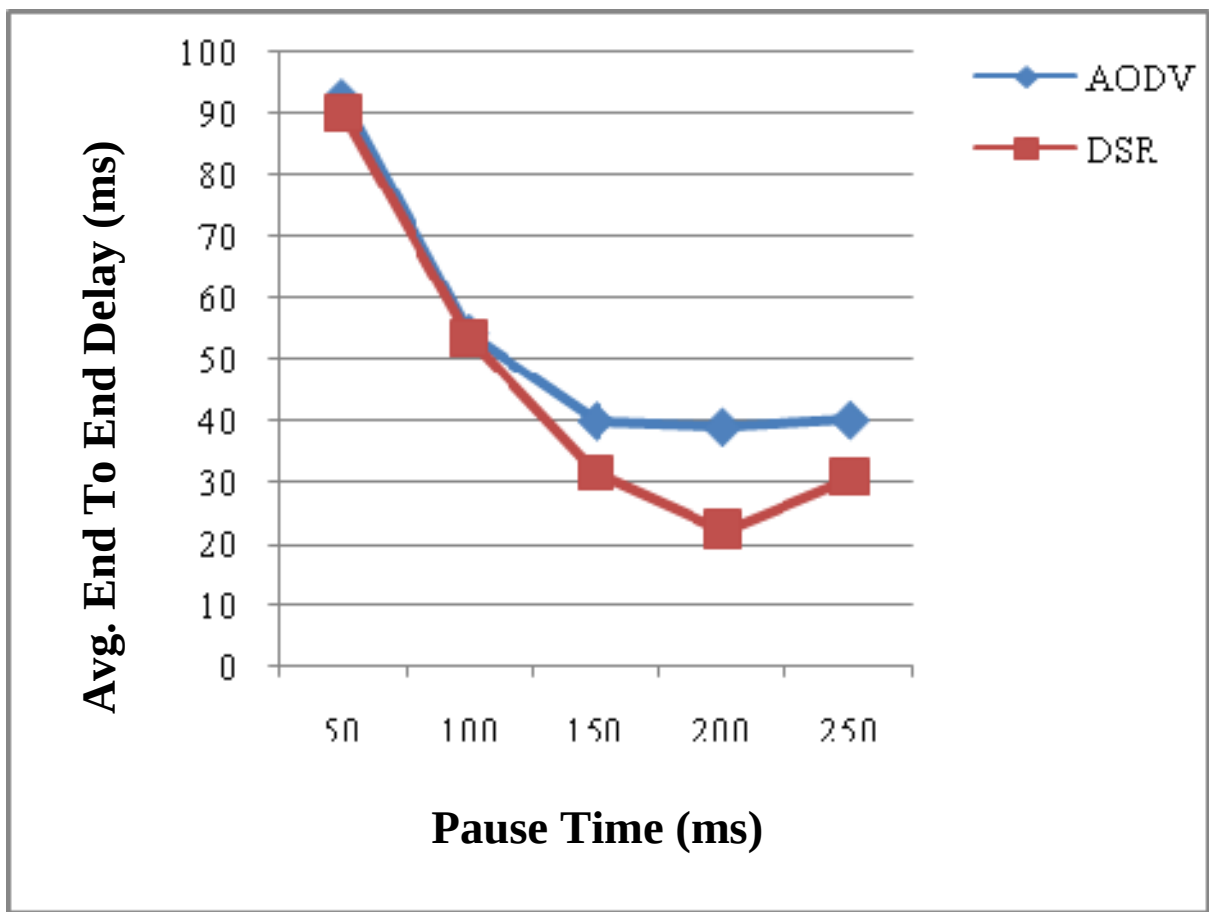

Fig 23: Avg.E-2-E delay of 120 nodes using CBR

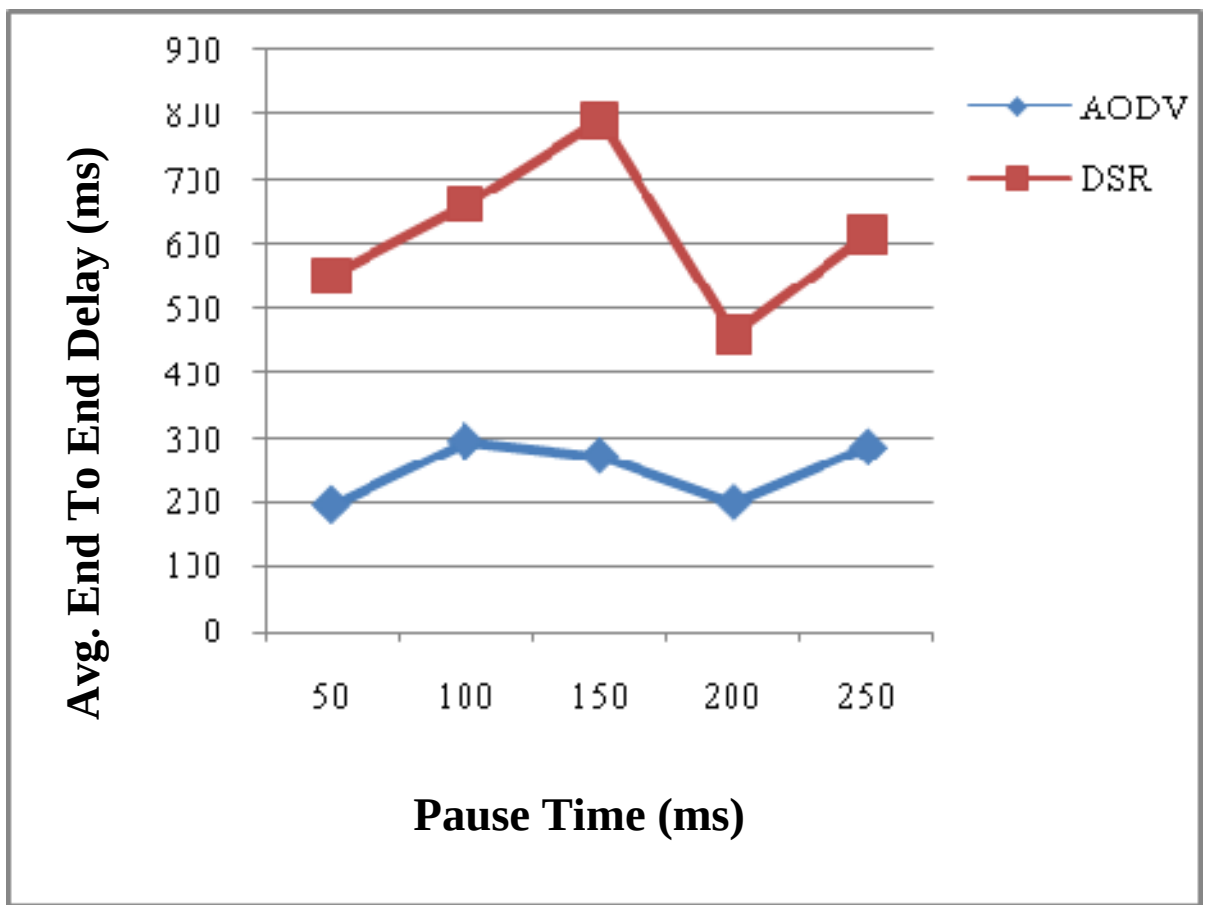

Fig 26: Avg.E-2-E delay of 150 nodes using TCP

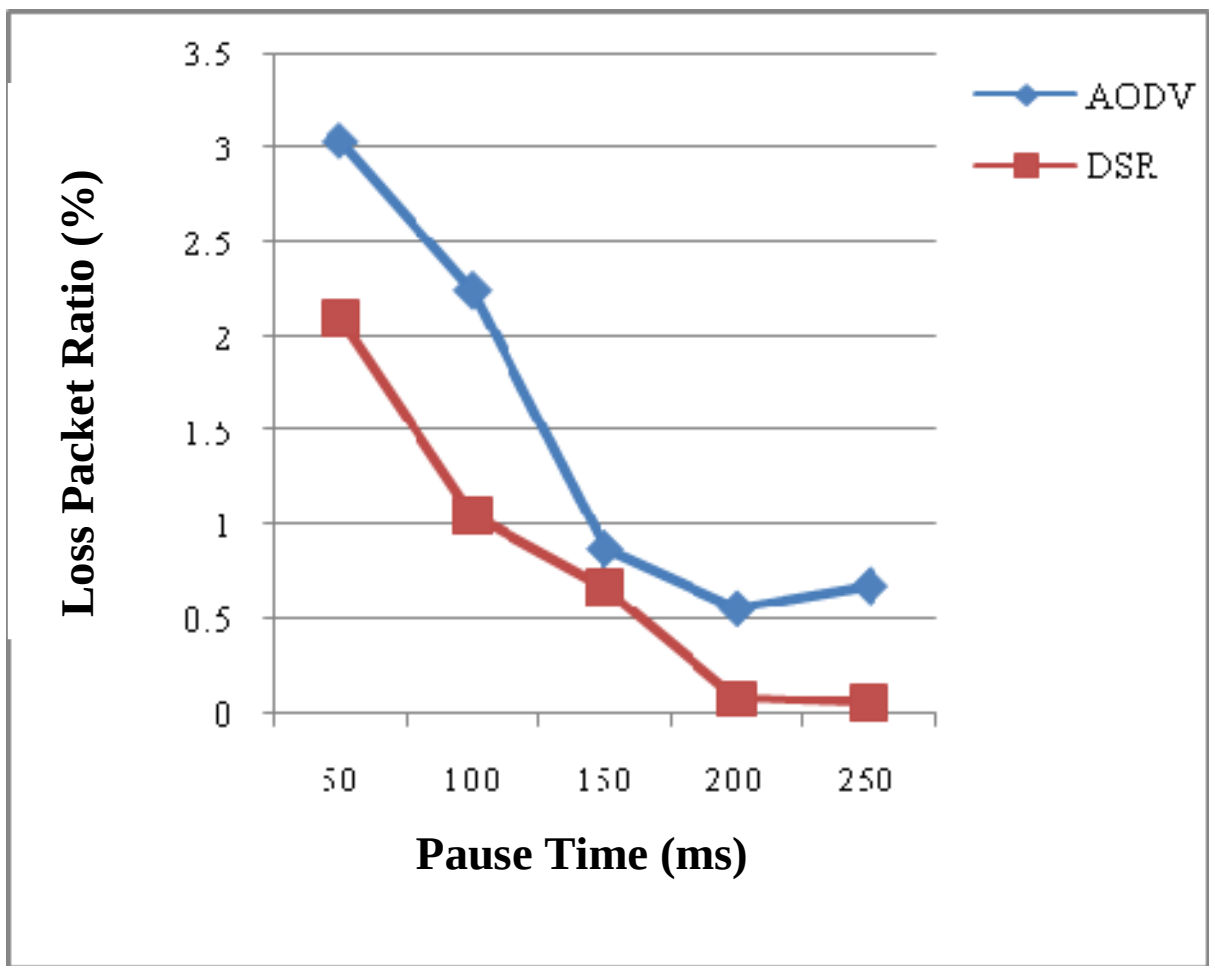

Fig 24: LPR of 120 nodes using CBR

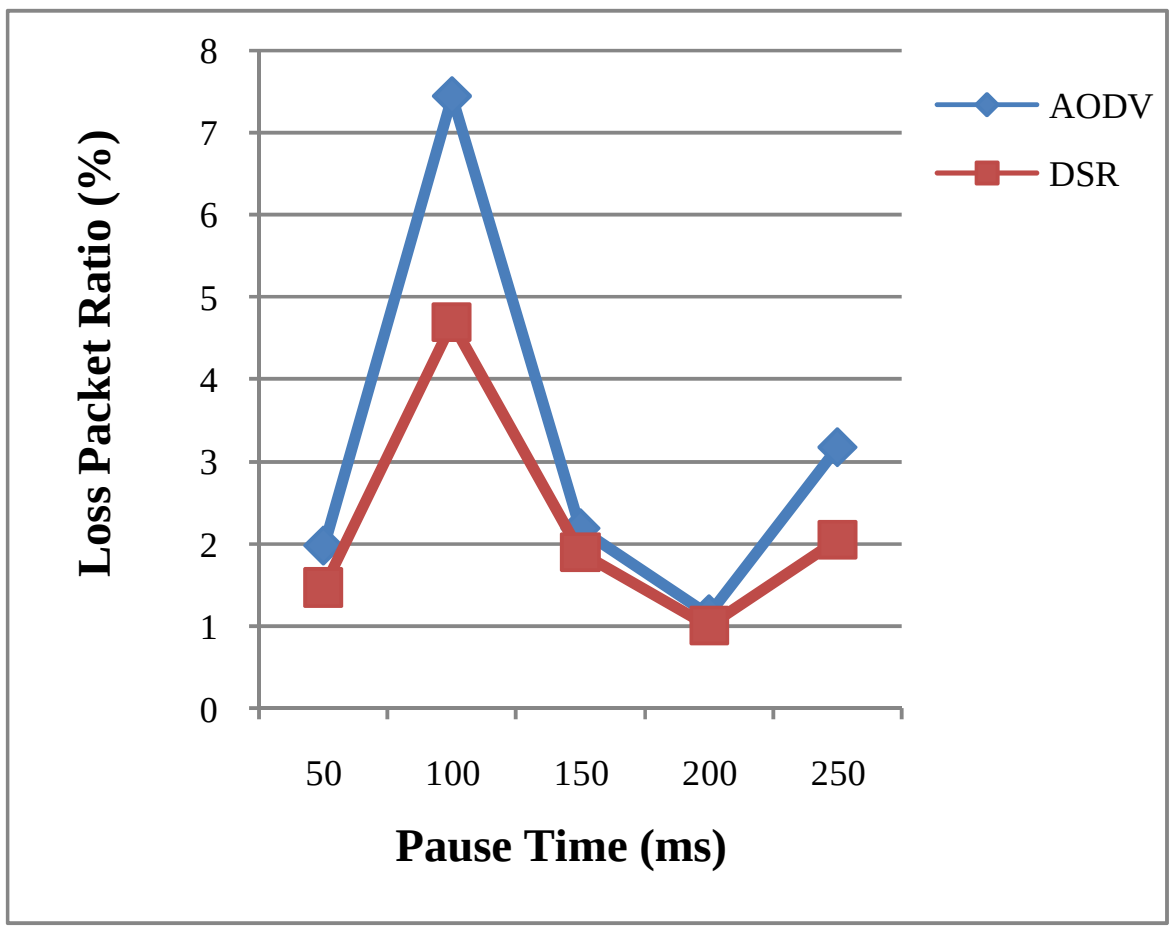

Fig 27: LPR of 150 nodes using TCP

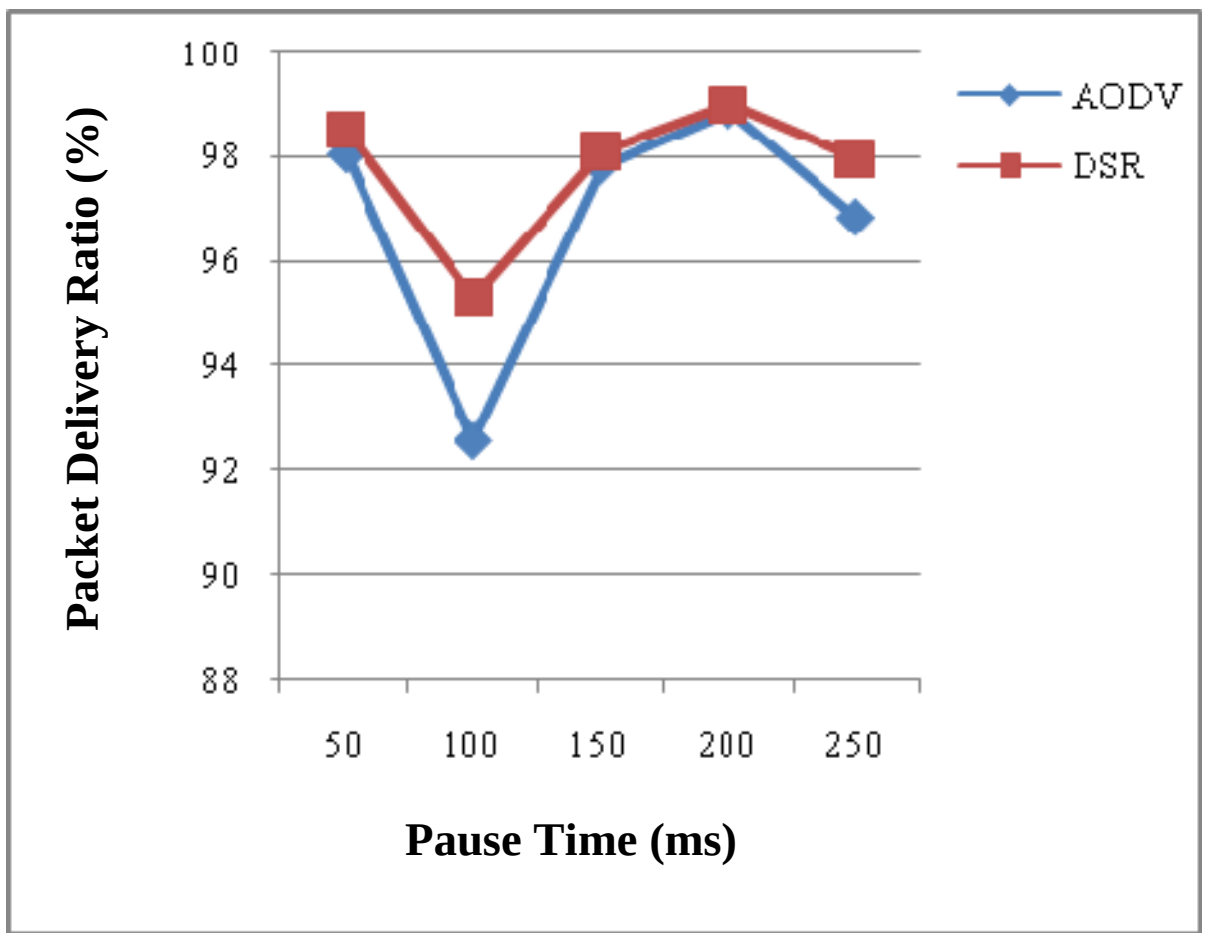

Fig 25: PDR of 150 nodes using TCP

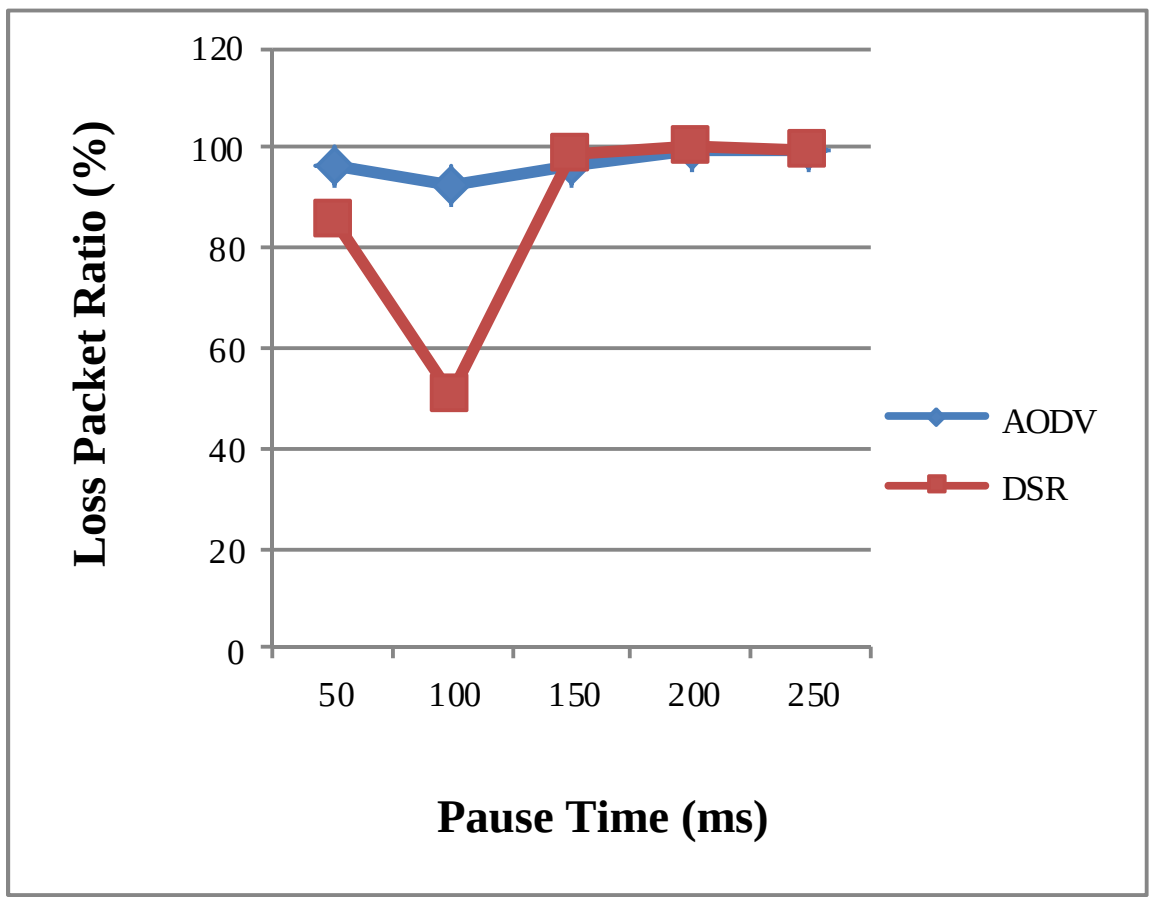

Fig 28: PDR of 150 nodes using CBR

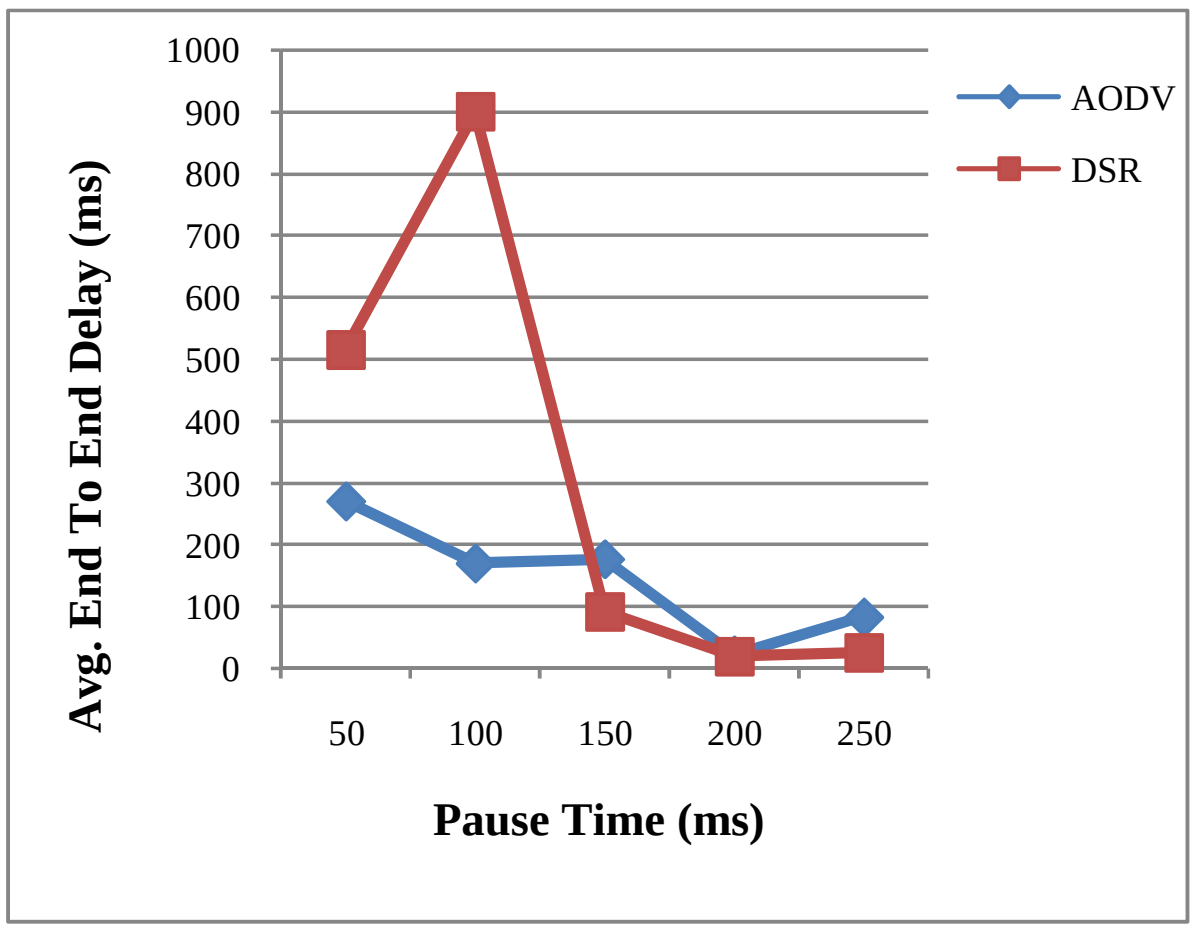

Fig 29: Avg.E-2-E delay of 150 nodes using CBR

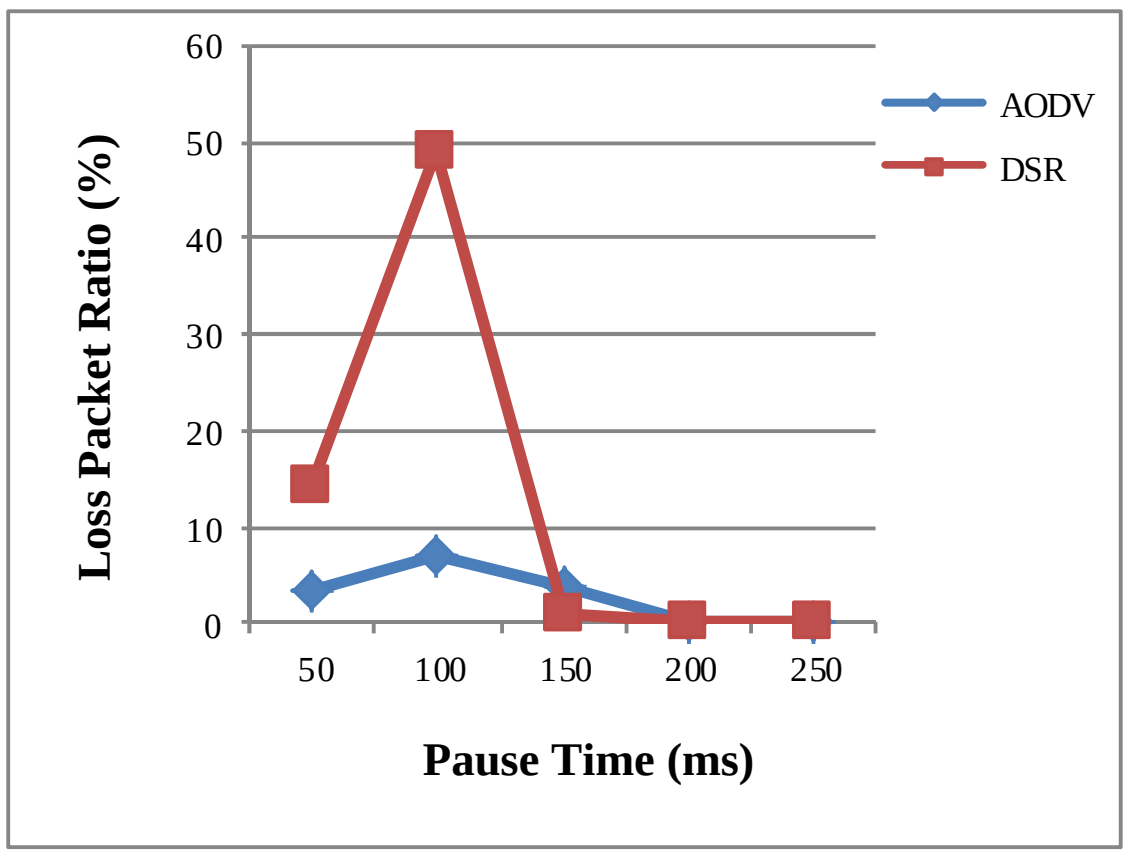

Fig 30: LPR of 150 nodes using CBR

| ***Node Density*** | ***Packet Delivery Ratio*** | | | | ***Avg. End to End Delay*** | | | | ***Loss Packet Ratio*** | | | |
|---|---|---|---|---|---|---|---|---|---|---|---|---|
| | **TCP** | | **CBR** | | **TCP** | | **CBR** | | **TCP** | | **CBR** | |
| **Low Density** | **AODV** | **DSR** | **AODV** | **DSR** | **AODV** | **DSR** | **AODV** | **DSR** | **AODV** | **DSR** | **AODV** | **DSR** |
| LowPauseTime | Avg | High | High | High | Avg | High | Low | Low | High | Avg | Avg | Low |
| Avg PauseTime | High | High | High | High | Avg | High | Low | Low | Avg | Low | Low | Low |
| HighPauseTime | High | High | High | High | High | High | Low | Low | Avg | Low | Low | Low |
| **Avg. Density** | | | | | | | | | | | | |
| LowPauseTime | Avg | Avg | Avg | High | High | High | Low | Low | High | High | High | Avg |
| Avg PauseTime | Avg | High | High | Avg | Avg | High | Low | Low | High | Avg | Low | Avg |
| HighPauseTime | Low | High | High | High | Avg | High | Low | Low | High | Avg | Low | Low |
| **High Density** | | | | | | | | | | | | |
| LowPauseTime | High | High | Avg | Low | Avg | High | Avg | High | Avg | Avg | High | High |
| Avg PauseTime | Avg | High | Avg | High | Avg | High | Avg | Low | High | Avg | High | Low |
| HighPauseTime | Avg | Avg | High | High | Avg | High | Low | Low | High | High | Low | Low |

## 7. Conclusion

This paper illustrates the differences between AODV and DSR based on TCP and CBR connection with various network parameters. In our analytical table we have given our decision based on the graph. This will definitely help to understand the performance of these two routing protocol.
The performance of these two routing protocol shows some differences in low and high node density.

From our experimental analysis we can conclude that in low density with low pause time the packet delivery ratio (PDR) of CBR connection for both routing protocols is high but for TCP connection average for AODV and high for DSR.

In that scenario average end to end delay (E-To-E) is low for both protocols using CBR connection. But for TCP connection average for AODV and high for DSR .The loss packet ratio for TCP connection is high for AODV and average for DSR .Average for AODV and low for DSR using CBR connection. If the pause time is high the PDR for both routing protocols using TCP and CBR is high .E-To-E for TCP is high and low for CBR connection for both routing protocols. LPR of DSR using TCP and CBR connection is low. But for AODV using TCP it is average and low for CBR connection.

In high density with low pause time, PDR for both protocols is high for TCP connection. In CBR connection, it is average for AODV and low for DSR. E-To-E for AODV using TCP and CBR connection is average but it is high for DSR .If the pause time is high the PDR for AODV and DSR using CBR is high but using TCP average for both protocols .E-To-E using CBR is low for both routing protocols but using TCP average for AODV and high for DSR. LPR using TCP connection is high and low using CBR for AODV and DSR.

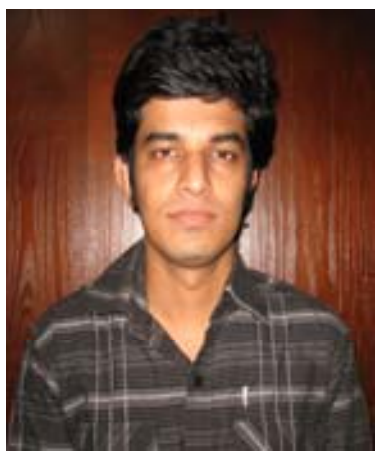

**Bijan Paul** is a B.Sc. student in the Dept. of Computer Science & Engineering, Shahjalal University of Science & Technology, Bangladesh. His research interests includes VANET, Routing protocols, Wireless computing.

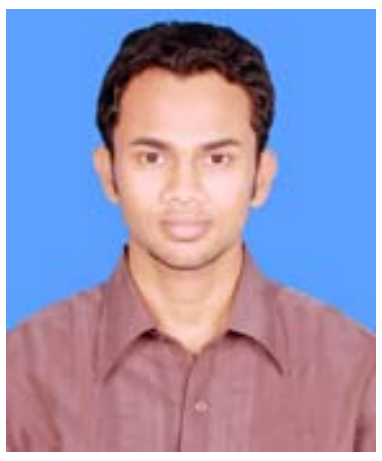

**Md. Ibrahim** is a B. Sc. student in the Dept. of Computer Science & Engineering, Shahjalal University of Science & Technology, Bangladesh. His research interest includes VANET, Routing protocols, Wireless Computing.

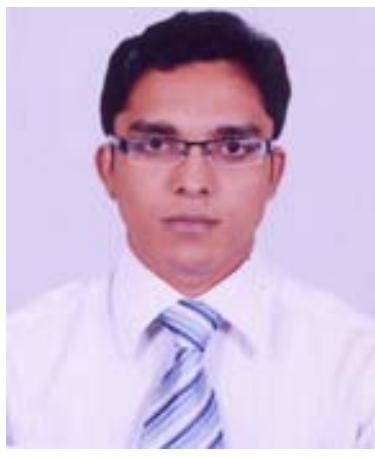

**Md. Abu Naser Bikas,** obtained his B. Sc. Degree in Computer Science & Engineering from Shahjalal University of Science & Technology, Bangladesh. Currently, he is a Lecturer in Computer Science & Engineering at the same University. His research interests include VANET, Network Security, Intrusion Detection and Intrusion Prevention, Bangla OCR, Wireless Ad-Hoc Networks, and Grid Computing.